\documentclass[a4paper,11pt]{article}
\pdfoutput=1 

\usepackage{jheppub} 

\usepackage[T1]{fontenc} 

\usepackage{amsmath,amsfonts,amsbsy,amssymb,array,accents,dsfont}
\usepackage{enumerate,latexsym,graphicx}
\usepackage{xcolor,tikz}
\usepackage{tcolorbox}

\usepackage{subfigure}
\usepackage{stmaryrd}
\usepackage{dsfont}

\def\cA{\mathcal {A}}  \def\cC{\mathcal {C}}
  
\def\cG{\mathcal {G}} \def\cH{\mathcal {H}} 
\def\cJ{\mathcal {J}} \def\cK{\mathcal {K}} 
\def\cM{\mathcal {M}}

\def\cV{\mathcal {V}} \def\cW{\mathcal {W}} 
 \def\cZ{\mathcal {Z}}

\newcommand{\beq}{\begin{equation}}
\newcommand{\eeq}{\end{equation}}
\newcommand{\bea}{\begin{eqnarray}}
\newcommand{\eea}{\end{eqnarray}}

\newcommand{\vep}{\varepsilon}

\newcommand{\der}{\partial}

\newcommand{\nn}{\nonumber}

\newcommand{\of}[1]{\left(#1\right)}
\newcommand{\off}[1]{\left[#1\right]}
\newcommand{\offf}[1]{\left\{#1\right\}}

\usepackage{braket}
\usepackage{tikz}
\usetikzlibrary{matrix,calc,positioning,decorations.markings,decorations.pathmorphing,decorations.pathreplacing}
\usetikzlibrary{arrows,cd}
\usepackage{bbding}
\usetikzlibrary{positioning}
\tikzset{>=stealth}

\newcommand{\psisl}{\psi}
\newcommand{\psisu}{\chi}

\newcommand{\del}{\partial}

\newcommand{\qqquad}{\;, \quad\qquad}  

\newcommand{\RR}{\mathbb{R}}
\newcommand{\ZZ}{\mathbb{Z}}
\newcommand{\NN}{\mathbb{N}}

\newcommand{\Nn}{\mathcal{N}}

\newcommand{\Uu}{\mathcal{U}}
\newcommand{\Vv}{\mathcal{V}}

\newcommand{\R}{\ensuremath{\mathbb{R}}}

\newcommand{\mb}{\bar{m}}

\newcommand{\xb}{\bar{x}}

\newcommand{\w}{\omega}

\usepackage{color}

\usepackage[normalem]{ulem}  

\definecolor{darkred}{rgb}{0.6,0,0}
\definecolor{darkblue}{rgb}{0,0,0.6}

\newcommand\p{\partial}

\newcommand{\be}{\begin{equation}}
\newcommand{\ee}{\end{equation}}

\usepackage[bbgreekl]{mathbbol}
\usepackage{amsfonts}
\DeclareSymbolFontAlphabet{\mathbb}{AMSb} 
\DeclareSymbolFontAlphabet{\mathbbl}{bbold} 

\title{\boldmath Spectral flow and string correlators in AdS$_3\times S^3 \times T^4$}

\author[a,b
]{Sergio Iguri,}
\author[c]{Nicolas Kovensky}
\author[d]{and Juli\'an H.~Toro}

\affiliation[a]{Instituto de Astronomía y Física del Espacio (IAFE) - CONICET and Facultad de Ciencias Exactas y Naturales, Universidad de Buenos Aires, Ciudad Universitaria, 1428 Buenos Aires, Argentina.}
\affiliation[b]{Mathematics with Computer Science Program, Guangdong Technion - Israel Institute of Technology, 515063 Shantou, Guangdong, People's Republic of China.}
\affiliation[c]{Institut de Physique Th\'eorique, Universit\'e Paris Saclay, CEA, CNRS, Orme des Merisiers, 91191 Gif-sur-Yvette CEDEX, France.}
\affiliation[d]{Instituto de Investigaciones Matemáticas Luis A. Santaló, CONICET - Universidad de Buenos Aires, Ciudad Universitaria, 1428 Buenos Aires, Argentina.}
\emailAdd{siguri@iafe.uba.ar}
\emailAdd{nicolas.kovensky@ipht.fr}
\emailAdd{jtoro@dm.uba.ar}

\abstract{We consider three-point correlation functions for superstrings propagating in AdS$_3\times S^3 \times T^4$. In the RNS formalism, these generically involve correlators with current insertions. When vertex operators with non-trivial spectral flow charges are present, their complicated OPEs with the currents imply that standard methods can not be used to compute such correlators. Here we develop techniques for computing all $m$-basis correlators of the supersymmetric model. We then show how, in some cases, these results can be translated to the $x$-basis. We obtain a new family of holographic three-point functions involving spacetime chiral primaries living in spectrally flowed sectors of the worldsheet CFT. These match precisely the predictions from the holographic dual at the symmetric product orbifold point. Finally, we also consider long strings and compute the probability amplitude associated with the process describing the emission/absorption of fundamental string quanta. 
}

\begin{document} 
\maketitle

\section{Introduction}

The propagation of superstrings in AdS$_3\times S^3 \times T^4$ (or $K3$) backgrounds has become one of the most fruitful frameworks for exploring the AdS/CFT correspondence. Unlike for the case of AdS$_5\times S^5$, perhaps the other prototypical model for studying holography, identifying the corresponding dual quantum field theory has been an elusive task\footnote{In fact a precise albeit perturbative definition has been provided only recently in \cite{Eberhardt:2021vsx}, a proposal that was further tested in \cite{Dei:2022pkr}. }. Nevertheless, the AdS$_3$ case outperforms the higher-dimensional one in that it provides a scenario in which one can describe a wide range of exciting phenomena beyond the supergravity limit. Indeed, one has access to a worldsheet description which is, in principle, exactly solvable \cite{Teschner:1999ug,Giveon:1998ns,Kutasov:1999xu,Maldacena:2000hw,Maldacena:2000kv,Maldacena:2001km}. The relevance of this model has recently increased due not only to multiple advances in the computation of string correlation functions \cite{Dei:2021xgh,Dei:2021yom,Iguri:2022eat} and the construction of the associated tensionless limit \cite{Giribet:2018ada,Gaberdiel:2018rqv,Eberhardt:2018ouy}, but also because of the role they play in the microscopic description of black holes \cite{Martinec:2017ztd,Martinec:2018nco,Martinec:2019wzw,Martinec:2020gkv,Balthazar:2021xeh}. The study of string correlation functions in this context has revealed crucial information about fundamental aspects of holography \cite{Eberhardt:2019ywk,Eberhardt:2020bgq} and black hole phenomena \cite{Jafferis:2021ywg,Bufalini:2021ndn,Bufalini:2022wyp,Bufalini:2022wzu,Bena:2022rna}, as well as single-trace $T\bar{T}$ and $J \bar{T}$ deformations of 2d CFTs and holography beyond AdS \cite{Giveon:2017myj,Asrat:2017tzd,Chakraborty:2018vja,Chakraborty:2019mdf}, and even condensed matter \cite{Polchinski:2012nh}. 

Although the determination of the spectrum and the calculation of correlators in the $S^3$ bosonic sector can be considered canonical, being the associated sector of the worldsheet CFT a conformal rational model, the AdS$_3$ sector presents several unusual features, mainly due to the fact that the underlying conformal field theory is a Wess-Zumino-Witten (WZW) model built upon a non-compact Lorentzian target space, i.e.~the universal cover of SL(2,$\mathbb R$). A consistent set of physical states necessarily involves the non-trivial action of the so-called spectral flow automorphisms \cite{Maldacena:2000hw}. Consistency of the conjectured spectrum was strengthened after determining the associated partition function \cite{Maldacena:2000kv}. As opposed to the rational case, spectral flow gives rise to representations that are inequivalent to the canonical ones, which encompass, in particular, a continuum of long string configurations, namely the states describing strings that can reach the asymptotic boundary while remaining finite in energy. 

Spectral flow also plays a crucial role in the short-string sector of the theory. Focusing on the supersymmetric version of the AdS$_3\times S^3 \times T^4$ model, when the AdS scale is larger than the string scale this sector contains the operators that are dual to chiral primaries of the spacetime theory. One can think of these operators in terms of the symmetric orbifold CFT Sym$_{N}\left(T^4\right)$, which lives in the same moduli space. Here $N=n_1 n_5$ is identified in terms of the number of NS5-branes and fundamental strings sourcing the near-horizon geometry. The unflowed sector of the worldsheet theory describes a subset of these operators, i.e.~those with low-lying spacetime weights $H< n_5/2$. As it was shown in \cite{Argurio:2000tb,Dabholkar:2007ey}, (almost) all of the remaining chiral primaries correspond to spectrally flowed operators. The precise form of the corresponding vertex operators was provided in \cite{Giribet:2007wp}.

The systematic computation of correlation functions for the bosonic string in pure AdS$_3$ was initiated in \cite{Maldacena:2001km}. Several of the papers that continued this line of work \cite{Giribet:2000fy,Giribet:2001ft,Hofman:2004ny,Giribet:2005ix,Ribault:2005ms,Iguri:2007af,Iguri:2009cf,Giribet:2011xf,Giribet:2015oiy} dealt with the calculation of correlators in the basis in which the Cartan current is diagonal. This so-called $m$-basis is best suited for performing the spectral flow operation. Correlation functions in this representation have proven to be useful for describing string interactions both in AdS$_3$ and in related cosets \cite{Martinec:2018nco,Bufalini:2022wyp,Bufalini:2022wzu} from a worldsheet perspective.
Other papers deal with \textit{alternative} bases, such as the so-called $t$-basis \cite{Ribault:2009ui} or the $x$-basis \cite{Teschner:1999ug,Minces:2005nb,Dabholkar:2007ey,Giribet:2007wp,Cagnacci:2013ufa,Dei:2021xgh}. The latter is best suited when it comes to holographic applications of the theory, as the continuous parameter $x$ is identified with the (holomorphic) coordinate on the AdS$_3$ boundary where the holographic CFT is defined. As $m$ and $x$ are conjugate variables, integrated vertex operators in the $x$-basis are local in spacetime, and they can be expressed as a sum over (an infinite number of) $m$-basis fields, i.e.~their spacetime Virasoro modes.

Vertex operators with non-trivial spectral flow charges are not affine primaries. Their complicated OPEs with the affine currents, together with the fact that the spectral flow charge is not a conserved quantity, render the computation of correlation functions involving these operators rather complicated. In the bosonic theory, a number of papers \cite{Minces:2005nb,Cagnacci:2013ufa} have managed to compute certain subfamilies of their $x$-basis three-point functions by using and further developing some techniques introduced in \cite{Maldacena:2001km}. More recently, an interesting novel approach was developed in \cite{Dei:2021xgh,Dei:2021yom,Dei:2022pkr}, based on a set of Ward identities arising in spectrally flowed sectors, leading to a number of conjectures regarding the analytic structure of spectrally-flowed correlators for more general configurations. The relation between these complementary methods was recently clarified in \cite{Iguri:2022eat}. 

We will be interested in three-point functions of spectrally-flowed operators in the supersymmetric AdS$_3\times S^3 \times T^4$ model. Some of these, involving short strings and conserving the total amount of spectral flow, were discussed in \cite{Giribet:2007wp}. Two major new issues that, unlike the bosonic cases, must be carefully treated, emerge in this context. On the one hand, the collusion of AdS$_3$ and $S^3$ strongly impacts the definition of the physical spectrum. Both the GSO projection and BRST invariance constrain the consistent configurations contributing to the supersymmetric model. They require, in particular, the introduction of spectral flow on the $3$-sphere when dealing with discrete states
. On the other hand, the picture-changing procedure one needs to implement to cancel the background ghost charge in the RNS formalism renders   the computation of bosonic correlators involving affine current insertions necessary. These have not been determined so far, as they cannot be derived in the usual way due to the peculiar properties of the spectrally flowed states, which produce many unknown terms in their OPEs with the currents. They are moreover required in order to complete the study initiated in \cite{Giribet:2007wp} for three-point functions of spacetime chiral primaries from the worldsheet CFT. These constitute protected quantities \cite{Baggio:2012rr}, which can and should be matched with the exact results obtained from the symmetric orbifold CFT\footnote{This matching was recently completed in \cite{Gaberdiel:2022oeu} for the somewhat special case of $n_5=1$ \cite{Eberhardt:2018ouy}, whose worldsheet theory  must be formulated in the so-called hybrid formalism \cite{Berkovits:1999im}.} \cite{Jevicki:1998bm,Lunin:2000yv,Dabholkar:2007ey}. 

In this paper, we address an important gap in the literature by providing explicit expressions for these correlation functions. We first develop a method for dealing with bosonic correlators involving the relevant affine current insertions in the $m$-basis, based on \cite{Maldacena:2001km,McElgin:2015eho}. The computation is done for arbitrary spectral flow assignments, that is, we compute all descendant three-point functions of the form  
\beq
\braket{\off{j^{a,\omega_1} V^{\omega_1}_{h_1, m_1}}V^{\omega_2}_{h_2, m_2}V^{\omega_3}_{h_3, m_3}},
\eeq
where $j^{a}$ and $V$ are the bosonic currents and spectrally flowed primaries, respectively. As already mentioned, these expressions are not only relevant from a pure AdS$_3$ perspective, but they can be useful in the context of other related coset models. Next, in order to deduce the corresponding results for operators in the spacetime representation, we follow a path similar to that introduced in \cite{Maldacena:2001km} and further developed in \cite{Cagnacci:2013ufa}. Although we are not able to obtain the $x$-basis correlators in full generality, as was the case in \cite{Cagnacci:2013ufa}, we compute this kind of three-point functions for a wide range of spectral flow assignments. 
More precisely, we focus on the NS-NS sector of the worldsheet theory, and compute 
\beq
\braket{\cV^{(0)}_{h_1}(x_1)\cV^{\omega_2}_{h_2}(x_2)\cV^{\omega_3}_{h_3}(x_3)},
\eeq
where $\cV^{\w}$ are the supersymmetric vertex operators and $h$ are their spacetime weights, while the superscript ``(0)'' stands for the ghost picture-zero version of the corresponding unflowed operator.  

We then show that the precise matching with the holographic CFT results holds for the associated families of chiral primary correlators as well, thus extending the analysis of \cite{Giribet:2007wp}.
Although somewhat restrictive, our results further allow us to explore, for the first time, how processes, where the background emits/absorbs a unit of fundamental string charge take place within the supersymmetric model with $n_5>1$. 

The paper is organized as follows. In Section 2, after reviewing some basics about superstring theory in AdS$_3\times S^3 \times T^4$, we re-compute explicitly all unflowed bosonic three-point functions involving descendant insertions, both in the $m$ and in the $x$ basis, obtaining the unflowed three-point functions in a way that is well suited for the generalization to the spectrally flowed cases. In Section 3 we discuss in detail the vertex algebra in spectrally flowed sectors and the so-called series identification. We define field operators for both short and long strings and their ghost picture-zero versions. We also compute the corresponding two-point functions. Then, in Section 4, we extend the computation of three-point correlators in the $m$ basis to arbitrary spectrally flowed insertions by introducing a method for dealing with bosonic correlation functions with descendant insertions. We also compute three-point correlators in the $x$ basis. Finally, in Section 5, we present our concluding remarks.

\section{Superstring theory on AdS$_3\times S^3\times T^4$}

We start by briefly reviewing the relevant aspects of superstring theory on AdS$_3\times S^3\times T^4$, the near-horizon region of the background sourced by $n_5$ NS5-branes and $n_1$ fundamental strings. The main building block of the worldsheet theory is given by the SL(2,$\R$)-WZW model introduced in  \cite{Teschner:1999ug,Maldacena:2000hw,Maldacena:2000kv,Maldacena:2001km}. The supersymmetric extension was investigated in \cite{Giveon:1998ns,Kutasov:1999xu,Argurio:2000tb}, while the corresponding extremal correlators were studied in  \cite{Dabholkar:2007ey,Gaberdiel:2007vu,Giribet:2007wp}. 

\subsection{Basic definitions}
The propagation of strings in AdS$_3\times S^3\times T^4$ is described in terms of a product of supersymmetric SL(2,$\R$) and SU(2) affine algebras at level $n_5$, whose currents and fermions will be denoted as $J^{a}$, $\psi^a$, $K^\alpha$ and $\chi^\alpha$, respectively, with $a,\alpha=0,1,2$. They satisfy the OPEs 
\begin{subequations}
\bea
&J^a(z)J^b(w) &\sim \frac{\frac{n_5}{2}\eta^{ab}}{(z-w)^2}+\frac{i\epsilon^{ab}{}_{c}J^c(w)}{z-w},\\
&J^a(z)\psi^b(w)&\sim \frac{i\epsilon^{ab}{}_{c}\psi^c(w)}{z-w},\\
&\psi^a(z)\psi^b(w)&\sim \frac{\frac{n_5}{2}\eta^{ab}}{z-w}\label{SL fermion OPE},
\eea
\end{subequations}
and 
\begin{subequations}
\bea
&K^\alpha(z)K^\beta(w) &\sim \frac{\frac{n_5}{2}\delta^{\alpha\beta}}{(z-w)^2}+\frac{i\epsilon^{\alpha\beta}{}_{\gamma}K^\gamma(w)}{z-w},\\
&K^\alpha(z)\chi^\beta(w)&\sim \frac{i\epsilon^{\alpha\beta}{}_{\gamma}\chi^\gamma(w)}{z-w},\\
&\chi^\alpha(z)\chi^\beta(w)&\sim \frac{\frac{n_5}{2}\delta^{\alpha\beta}}{z-w},
\eea
\end{subequations}
where $\epsilon^{012}=1$ $\eta^{ab} = \eta_{ab} = (-++)$, and $\delta^{\alpha\beta}=\delta_{\alpha\beta} = (+++)$. We will mostly use the ladder operators $J^\pm = J^1 \pm i J^2$, and similarly for $K^\pm$, $\psi^\pm$ and $\chi^\pm$. The supersymmetric currents split as 
\beq
J^a = j^a + \hat{j}^a\qqquad
K^\alpha = k^\alpha + \hat{k}^\alpha,
\eeq
where $j^a$ and $k^\alpha$ generate bosonic affine algebras SL(2,$\R$)$_{k}$ and SU(2)$_{k'}$ with shifted levels with levels $k = n_5 + 2$ and $k' = n_5-2$, while 
\beq 
\hat{j}^a = -\frac{i}{n_5}\epsilon^{a}{}_{bc}\psi^b \psi^c\qqquad 
\hat{k}^\alpha = -\frac{i}{n_5}\epsilon^{\alpha}{}_{\beta\gamma}\chi^\beta \chi^\gamma
\eeq
generate fermionic SL(2,$\R$)$_{-2}$ and SU(2)$_{2}$ algebras. Finally, we also have the free bosons and fermions associated with the $T^4$ directions, which will be denoted as $Y^i$ and $\lambda^i$, respectively, with $i=6,\dots,9$.


It will also be useful to use the bosonized forms for the SL(2,$\R$) and SU(2) fermions. We consider canonically normalized bosonic fields $H_I$, with $I=1,\dots 5$, and introduce 
\begin{equation}
        \hat{H}_I = H_I + \pi \sum_{J<I} N_J \qqquad N_J \equiv \oint i\del H_J \, , 
\end{equation}
where the number operators $N_I$ are introduced in order to keep track of the cocycle factors, with 
\begin{equation}
        e^{i a \hat{H}_I} e^{i b \hat{H}_J} = e^{i b \hat{H}_J} e^{i a \hat{H}_I} \, e^{i \pi a b} \qqquad \text{if} \quad I > J \, . \label{HIAdS3}
\end{equation}
The bosonization of $\psi^a$ and $\chi^\alpha $ then reads 
\begin{subequations}
  \begin{gather}
        i\del \hat{H}_1 = \frac{1}{n_5} \, \psisl^+\psisl^- \qqquad   i \del \hat{H}_2 =  \frac{1}{n_5} \, \psisu^+\psisu^-  \qqquad
	 i \del \hat{H}_3 = \frac{2}{n_5} \, \psisl^0\psisu^0  \, ,  
    \\  i\del \hat{H}_4 = i\lambda^6 \lambda^7 \qqquad  i\del \hat{H}_5 = i\lambda^8 \lambda^9 \, ,
	\end{gather}
\end{subequations}  
where $ \hat{H}_I^\dagger = \hat{H}_I$ for $I \ne 3 $ and $ \hat{H}_3^\dagger = - \hat{H}_3 $. Conversely, 
\begin{subequations}
\begin{gather}
	 \psisl^\pm = \sqrt{n_5} \, e^{\pm i\hat{H}_1} 
	 \, , \quad 
	\psisu^\pm = \sqrt{n_5} \, e^{\pm i \hat{H}_2} 
	 \, , \quad
	 \lambda^{6} \pm i \lambda^7 = e^{\pm i \hat{H}_4} 
	 \, , \quad 
	 \lambda^{8} \pm i \lambda^9 = e^{\pm i \hat{H}_5} \, ,  \label{psiH1xiH2} \\
	 \qquad \psisl^0 = \frac{\sqrt{n_5}}{2} \, \left(e^{i \hat{H}_3} - e^{-i\hat{H}_3} \right)  \qqquad \psisu^0 =  \frac{\sqrt{n_5}}{2}  \, \left(e^{i \hat{H}_3} + e^{-i\hat{H}_3} \right) \, .
\end{gather}
\end{subequations}
In the remainder of the paper we will mostly omit the hats and explicitly include the phase factors only if necessary.


The stress tensor and the supercurrent of the matter sector of the worldsheet CFT derived from the Sugawara construction are 
\begin{eqnarray}
    T &=& \frac{1}{n_5} \left(j^a j_a - \psisl^a \der \psisl_a + 
    k^a k_a - \psisu^a \der \psisu_a 
    \right) + \frac{1}{2}
    \left(\der Y^i \der Y_i - \lambda^i \der \lambda_i\right),
    \label{TAdS3S3T4def}
    \\
    G &=& \frac{2}{n_5} \left(
    \psisl^a j_a - \frac{1}{3n_5} f_{abc} \psisl^a \psisl^b \psisl^c + 
    \psisu^a k_a - \frac{1}{3n_5} f'_{abc} \psisu^a \psisu^b \psisu^c
    \right) + i \:\lambda^i \der Y_i \, .
    \label{GAdS3S3T4def}
\end{eqnarray}
We also have the standard $bc$ and $\beta \gamma$ ghost systems, leading to the BRST charge 
\begin{equation}
\label{eq:BRSToperator}
    {\cal{Q}} = \oint dz \left[ c \left(T + T_{\beta\gamma}\right) - \gamma \, G + c(\der c) b - \frac{1}{4} b \gamma^2\right] \, .
\end{equation}
The $\beta\gamma$ system is also bosonized as 
\begin{align}
	\beta = e^{-\varphi} \del \xi  \qqquad \gamma = \eta \:\! e^{\varphi} \,,
\end{align}
where $\varphi(z) \varphi(w) \simeq - \ln(z-w)$ has background charge $2$, and $\xi(z)\eta(w) \sim (z-w)^{-1}$.
The spacetime supercharges can be written as 
\begin{equation}
    Q_\vep = \oint dz \, e^{-\varphi/2} S_\vep \qqquad   S_\vep = \exp \left(\frac{i}{2} 
    \sum_{I=1}^{5}\vep_I H_I\right),
    \label{supercharges}
\end{equation}
where $S_\vep$ are spin fields and $\vep_I=\pm 1$. These are constrained by $\vep_1 \vep_2 \vep_3 = \vep_4 \vep_5 = 1$ due to BRST-invariance and mutual locality, giving the 
supercharges of the spacetime $\Nn=(4,4)$ superconformal algebra. Moreover, the R-symmetry of the boundary theory is generated by the worldsheet SU(2) currents.

\subsection{Vertex operators in the unflowed sector}

Let us define the physical vertex operators. We will mostly follow the conventions of \cite{Dabholkar:2007ey,Gaberdiel:2007vu,Giribet:2007wp}. We also focus on the holomorphic part of the theory, omitting the anti-holomorphic dependence in most of the expressions below.   

Since bosonic and fermionic currents commute, vertex operators factorize into a product of bosonic primaries and free fermions. Let $V_h(x,z)$ be an SL(2,$\RR$)$_k$ primary field of weight 
\beq 
\Delta_h = -\frac{h(h-1)}{n_5}\, ,
\eeq
and spin $h$. It satisfies 
\beq
j^a(z) V_h(x,w) \sim - \frac{D^a_{x,h}V_h(x,w)}{z-w} \label{eq: SL2 bosonic OPE},
\eeq
with
\beq
D^-_{x,h} = \p_x, \qquad D^0_{x,h} = x\p_x + h, \qquad D^+_{x,h} = x^2\p_x + 2hx \label{zero modes}
\eeq
The fields $V_h(x,z)$ are defined in the so-called $x$-basis, the $x$ variable being identified holographically with the complex coordinate of the boundary theory \cite{Kutasov:1999xu}. Their spacetime modes then correspond to the $m$-basis operators
\beq
V_{hm}(z) = \int d^2x x^{h+m-1}\xb^{h+\mb-1} V_{h}(x,z). 
\label{Medellin}
\eeq
The bosonic currents OPEs with the fields $V_{hm}(z)$ are 
\beq
j^a(z) V_{hm}(w) \sim (m-a (h-1))\frac{V_{h, m+a}(w)}{z-w}
\qqquad a= 0,\pm 1\, .
\eeq
The relevant unflowed representations of the zero-mode algebra are
\begin{enumerate}
    \item[-] \textbf{Principal series discrete representation of lowest/highest-weight:} these are built by acting with $j^\pm_0$ on the state $\ket{h,\pm h}$ (created by $V_{h,\pm h}(0)$ acting on the vacuum), which is annihilated by $j^\mp_0$: 
    \beq
    \mathcal{D}^\pm_h =  \big\langle \ket{h,m}, m = \pm(h + n), n\in \NN \, \big\rangle.
    \eeq
    Note that for operators in ${\cal{D}}^\pm_h$ one can invert \eqref{Medellin}. Indeed, the (residues of the) poles located at $x=0$ give states in the highest-weight representations, namely 
    \beq
    V_{h}(x,z) = \sum_{m,\mb=h}^{\infty}  x^{-h-m} \bar{x}^{-h-\mb}V_{h,m}(z)\,.
    \label{Medellin2}
    \eeq
    Conversely, operators in the lowest-weight representation give the power-series expansion around $x = \infty$. In particular, we have 
    \begin{equation}
        V_{h,-h}(z) = V_h(x=0,z)\qqquad 
        V_{h, h}(z) = \lim_{x,\bar{x}\to \infty} |x|^{4h}V_h(x,z).
        \label{xtombasis}
    \end{equation}
    
    \item[-]\textbf{Principal continuous series:} These are given by
    \beq
    \mathcal{C}^\alpha_h = \big\langle \ket{h,\alpha,m},\alpha\in [0,1),h=\frac{1}{2}+is, s\in \RR, m = \alpha + n, n\in \ZZ \,\big\rangle.
    \eeq
\end{enumerate}
It was shown in \cite{Maldacena:2000hw} that the spectrum of the SL(2,$\R$)$_{k}$-WZW model is built out of continuous and lowest weight representations with 
\beq
\frac{1}{2}< h < \frac{k-1}{2}
\label{rangeSL2R}
\eeq
together with their spectrally flowed images, defined below. 
\medskip

The bosonic SU(2)$_{k'}$ WZW model has primary fields $W_{l,n}$ with $ n = -l,\dots,l$ and weight 
\beq 
\Delta_l = \frac{l(l+1)}{n_5},
\eeq
where the spin $l$ is bounded by \cite{Zamolodchikov:1986bd}
\beq 
0\leq l \leq \frac{k'}{2}.
\eeq
The OPEs with the currents $k^\alpha$ are given by  
\begin{subequations}
\bea
&k^0(z) W_{l,n}(w) &\sim n \frac{W_{l,n}(w)}{z-w},\\
&k^\pm(z) W_{l,n}(w) &\sim (l+1\pm n) \frac{W_{l,n\pm1}(w)}{z-w}.
\eea
\label{SU2opesMbasis}
\end{subequations}
Similarly to the SL(2,$\RR$) case, one can make use of the isospin variables $y$, with 
\beq 
W_l(y,z) = \sum_{n=-l}^{l} y^{l-n} W_{l,n}(z), 
\eeq
so that Eqs.~\eqref{SU2opesMbasis} read 
\beq
k^\alpha(z) W_{l}(y,w) \sim - \frac{P^\alpha_{y,l}W_{l}(y,w)}{z-w},
\eeq
where 
\beq
P^-_{y,l} = -\p_y\qqquad P^0_{y,l} = y\p_y -l \qqquad P^+_{y,l} = y^2\p_y -2ly. \label{Py}
\eeq

The properties of SL(2,$\R$) and SU(2) bosonic primary operators can be written compactly in the following form:
\begin{subequations}
\bea
&J(x_1,z_1) V_h(x_2,z_2) &\sim j(x_1,z_1) V_h(x_2,z_2) \sim \frac{1}{z_{12}}\off{x_{12}^2\p_{x_2} - 2h x_{12}}V_h(x_2,z_2), 
\label{JV OPE} \\
&K(y_1,z_1) W_l(y_2,z_2) &\sim k(y_1,z_1) W_l(y_2,z_2) \sim \frac{1}{z_{12}}\off{y_{12}^2\p_{y_2} + 2l y_{12}}W_l(y_2,z_2),
\eea
\end{subequations}
where $x_{12} = x_1 - x_2$, and we have introduced the currents 
\begin{subequations}
\bea
& J(x,z) &= -J^+(z) + 2x J^0(z) - x^2 J^-(z) = j(x,z) + \hat{j}(x,z), \\
&K(y,z) &= -K^+(z) + 2y K^0(z) + y^2 K^-(z) = k(y,z) + \hat{k}(y,z),
\eea
\end{subequations}
which satisfy 
\bea
&J(x_1,z_1)J(x_2,z_2) &\sim n_5\frac{x_{12}^2}{z_{12}^2}+\frac{1}{z_{12}}\off{x_{12}^2\p_{x_2} + 2x_{12}}J(x_2,z_2)\label{JJ OPE}\\
&K(y_1,z_1)K(y_2,z_2) &\sim -n_5\frac{y_{12}^2}{z_{12}^2}+\frac{1}{z_{12}}\off{y_{12}^2\p_{y_2} + 2y_{12}}K(y_2,z_2)
\eea
It will be useful to work similarly with the fermions 
\begin{subequations}
\bea
& \psi(x,z) &= -\psi^+(z) + 2x\psi^0(z) - x^2 \psi^-(z), \\
&\chi(y,z) &= - \chi^+(z) + 2y \chi^0(z) + y^2 \chi^-(z),
\eea
\end{subequations}
for which 
\begin{subequations}
\bea
&J(x_1,z_1) \psi(x_2,z_2) &\sim \hat{j}(x_1,z_1) \psi(x_2,z_2) \sim \frac{1}{z_{12}}\off{x_{12}^2\p_{x_2} + 2x_{12}}\psi(x_2,z_2), \\
&K(y_1,z_1) \chi(y_2,z_2) &\sim \hat{k}(y_1,z_1) \chi(y_2,z_2) \sim \frac{1}{z_{12}}\off{y_{12}^2\p_{y_2} + 2y_{12} }\chi(y_2,z_2),
\eea
\end{subequations}
as expected for fields with SL(2,$\R$) spin $h = -1$ and SU(2) spin $l=1$, respectively.


In the supersymmetric theory, unflowed vertex operators that belong to the continuous series representations are projected out by GSO due to their tachyonic nature. Physical vertex operators polarized in the SL(2,$\R$) and SU(2) directions belonging to the unflowed sector are given in \cite{Kutasov:1998zh,Dabholkar:2007ey,Gaberdiel:2007vu}. We focus on operators dual chiral primaries of the holographically dual CFT\footnote{There are additional families of physical operators in the NSNS sector which are not dual to chiral primaries. We will not discuss them in this paper due to the fact that their three-point functions are not protected by supersymmetry. In any case, it would be interesting to use the techniques developed here to compute the corresponding three-point functions in order to compare them with the correlators in the recent proposal of \cite{Eberhardt:2021vsx} for the holographic dual. Indeed, and as opposed to the usual symmetric orbifold theory considered for instance in \cite{Dabholkar:2007ey,Gaberdiel:2007vu}, this is conjectured to be the dual CFT at the same point in the moduli space where the worldsheet theory is defined, so that one should also be able to match non-protected quantities. }, namely
\begin{subequations}
\bea
&\cV_{h;l}(x,y) &= e^{-\varphi}\Phi_{h;l}(x,y)\psi(x),\\
&\cW_{h;l}(x,y) &= e^{-\varphi}\Phi_{h;l}(x,y)\chi(y).
\eea 
\end{subequations}
where $\Phi_{h;l}(x,y) = V_h(x) W_l(y)$ and the exponential of the free boson $\varphi$ comes from the ghost sector of the theory. The worldsheet dependence is omitted. These operators have definite (supersymmetric) spins $(H,L) = (h-1,l)$ and $(H,L) = (h,l+1)$, and represent massless excitations polarized in the AdS$_3$ and S$^3$ directions, respectively. Recall that one must also include fermion excitations in the anti-holomorphic sector.  The Virasoro conditions impose  
\beq
h(h-1) = l(l+1)\,,
\eeq
which is solved by setting $h = l+1$ (the second solution falls outside of the range \eqref{rangeSL2R}). Therefore, we will omit the subscript $l$ assuming this relation. In both cases, $H$ gives the corresponding spacetime weight and, since $H=L$ with $L$ the spacetime R-charge, these are chiral primary operators of the boundary theory. 

From now we will focus our attention on $\cV_h$. The treatment for $\cW_h$ is analogous. Using the relation \eqref{Medellin} it is possible to express $\cV_h$ in the $m$-basis. Explicitly, we have 
\beq
\cV_{h,m}(y) = \int d x \,  x^{H+m-1}\cV_h =  e^{-\varphi}\of{V_h \psi}_{h-1,m}W_{h-1}(y),
\eeq
with
\bea 
\of{V_h\psi}_{h-1,m} = -\psi^+ V_{h,m-1} + 2 \psi^0 V_{h,m} - \psi^- V_{h,m+1}, 
\eea 
and where we have once again ignored the anti-holomorphic sector.


\subsection{Two-point functions} 

The string theory two-point function factorizes into bosonic, fermionic and ghost contributions. The latter is given by $\braket{e^{-\varphi(z_1)}e^{-\varphi(z_2)}} = z_{12}^{-1}$. The SL(2,$\RR$) two-point function is given by \cite{Teschner:1997ft}
\beq
\braket{V_{h_1}(x_1,z_1)V_{h_2}(x_2,z_2)}=\frac{1}{z_{12}^{-2h_1(h_1-1)/n_5}}\off{\delta^2(x_1-x_2)\delta(h_1+h_2-1)+ B(h_1)\frac{\delta(h_1-h_2)}{x_{12}^{2h_1}}}\label{SL2 bosonic 2point}
\eeq
where \cite{Teschner:1999ug,Maldacena:2001km}
\beq
B(h) = - \frac{\nu^{-2h+1}}{\pi b^2}\gamma(1-b^2(2h-1))\qqquad \gamma(x) = \frac{\Gamma(x)}{\Gamma(1-\xb)}\qqquad b^2 = n_5^{-1},
\eeq
and $\nu$ is an arbitrary constant. 
As for the fermion propagator, we recall that the $x$- and $z$-dependence of the propagator is fixed by worldsheet and spacetime Ward identities.  $\psi(x,z)$ is a primary field of spin $h = -1$ and worldsheet weight $\Delta_\psi=1/2$ with respect to the $SL(2,\RR)_{-2}$ algebra generated by the currents $\hat{j}^a$. Making use of Eq.~\eqref{xtombasis}, we have 
\beq
\lim_{x_2\rightarrow \infty}x_2^{-2}\braket{\psi(x_1=0,z_1)\psi(x_2,z_2)}
= \braket{\psi^+(z_1)\psi^-(z_2)} = \frac{n_5}{z_{12}},
\eeq
such that 
\beq
\braket{\psi(x_1,z_1)\psi(x_2,z_2)} = n_5 \frac{x_{12}^2}{z_{12}},
\eeq
as expected. On the other hand, the SU(2) contributions read 
\beq 
\braket{W_{h_1-1}(y_1,z_1)W_{h_2-1}(y_2,z_2)} = \delta_{h_1,h_2}\frac{y_{12}^{2(h_1-1)}}{z_{12}^{2h_1(h_1-1)/n_5}}\, , 
\eeq
and 
\beq
\braket{\chi(y_1,z_1)\chi(y_2,z_2)} = -n_5 \frac{y_{12}^2}{z_{12}}.
\eeq
The full two-point functions then take the form 
\beq
\braket{\cV_{h_1} \cV_{h_2}} = \delta(h_1-h_2)n_5\frac{B(h_1)}{z_{12}^2}\of{\frac{y_{12}}{x_{12}}}^{2H_1},
\label{2ptVV}
\eeq
where $H_1 = h_1-1$,  and 
\beq
\braket{\cW_{h_1} \cW_{h_2}} = -\delta(h_1-h_2)n_5\frac{B(h_1)}{z_{12}^2}\of{\frac{y_{12}}{x_{12}}}^{2H_1}, 
\label{2ptWW}
\eeq
with $H_1 = h_1$. Note that the first term of the r.h.s of Eq. \eqref{SL2 bosonic 2point} does not contribute to the short string two-point functions. The factor $\delta(h_1+h_2-1)$ imposes $h_1+h_2=1$, which can not be archived if both $h_i$ are in the range \eqref{rangeSL2R}.

\subsection{Picture changing and three-point functions}\label{sec: unflowed 3point}

In order to compute string three-point functions in the NS sector of the theory it is necessary to obtain the ghost picture-zero version of the above vertex operators. As usual, the picture-changing operator is defined in terms of the total (matter) supercurrent $G$, so that  
\beq 
{\cal{O}}^{(0)}(z)= \lim_{w\rightarrow z}(e^{\varphi}G)(w) {\cal{O}}(z), 
\label{PictureChangingOp}
\eeq
where ${\cal{O}}$ stands for a generic vertex operator. For unflowed states, this was derived in  \cite{Dabholkar:2007ey,Gaberdiel:2007vu}, giving 
\begin{subequations}
\bea
&\cV_h^{(0)}(x,y) &= \of{(1-h)\hat{j}(x)+j(x)+\frac{2}{n_5}\psi(x)\chi_\alpha P^\alpha_{y,h-1}}\Phi_h(x,y) \, , \\
&\cW_h^{(0)}(x,y) &= \of{h\hat{k}(y)+k(y)+\frac{2}{n_5}\chi(y)\psi_a D^a_{x,h}}\Phi_h(x,y).
\eea
\label{PicZeroUnflowed}
\end{subequations}

Due to the second term appearing on the RHS of  Eqs.~\eqref{PicZeroUnflowed}, three-point correlators generically involve bosonic correlators with descendant insertions. More precisely, and focusing on the SL(2,$\R$) case, besides the usual primary correlators one needs  
\beq
\braket{\of{j V_{h_1}}(x_1) V_{h_2}(x_2) V_{h_3}(x_3)}. \label{no flow current}
\eeq
These correlators were computed in \cite{Dabholkar:2007ey,Gaberdiel:2007vu} directly in the $x$-basis from the OPEs between currents and primaries by means of the usual contour integration techniques. However, and as will be discussed below, this method is unavailable for vertex operators with arbitrary spectral flow charges since the corresponding OPEs contain many unknown terms \cite{Eberhardt:2019ywk} (except for the singly-flowed case \cite{Maldacena:2001km,Cardona:2009hk}). 

We now derive \eqref{no flow current} in an alternative way, best suited for the generalization to the spectrally flowed cases. Similar to what was done in \cite{Cagnacci:2013ufa} for primary correlators, the idea is to relate it with some $m$-basis correlator, and read out the corresponding structure constant. We first determine the $z$- and $x$-dependence the correlator. This is fixed by the action of the global currents. We know that $V_h$ is a primary field of spin $h$ and conformal weight $\Delta_h = -h(h-1)/n_5$. On the other hand, $\of{j V_h}$ is a descendant that has well-defined spin $h-1$ and weight $\Delta_h+1$. Indeed, using \eqref{JJ OPE} and \eqref{JV OPE} we have
\beq
j(x_1,z_1)\of{jV_h}(x_2,z_2) \sim \frac{kx_{12}^2}{z_{12}^2}V_h(x_2,z_2)+\frac{1}{z_{12}}\off{x_{12}^2\p_{x_2} - 2(h-1)x_{12}}\of{j V_h}(x_2,z_2) \label{1}.
\eeq
Hence,  
\beq
\braket{\of{j V_{h_1}}(x_1) V_{h_2}(x_2) V_{h_3}(x_3)}= C(h_i) \frac{x_{12}^{h_3-(h_1-1)-h_2}x_{23}^{(h_1-1)-h_2-h_3}x_{13}^{h_2-(h_1-1)-h_3}}{z_{12}^{\Delta_{h_1}+1+\Delta_{h_2}-\Delta_{h_3}}z_{23}^{\Delta_{h_2}+\Delta_{h_3}-\Delta_{h_1}-1}z_{13}^{\Delta_{h_1}+1+\Delta_{h_3}-\Delta_{h_2}}}\label{251}.
\eeq
The goal is to derive the relation between the structure constants $C(h_i)$ and those of the  primary three-point functions
\beq
\braket{V_{h_1}(x_1) V_{h_2}(x_2) V_{h_3}(x_3)} = C_H(h_i) \frac{x_{12}^{h_3-h_1-h_2}x_{23}^{h_1-h_2-h_3}x_{13}^{h_2-h_1-h_3}}{z_{12}^{\Delta_{h_1}+\Delta_{h_2}-\Delta_{h_3}}z_{23}^{\Delta_{h_2}+\Delta_{h_3}-\Delta_{h_1}}z_{13}^{\Delta_{h_1}+\Delta_{h_3}-\Delta_{h_2}}},
\eeq
derived in \cite{Maldacena:2001km,Teschner:1999ug} in terms of Barnes double Gamma functions. For this, we note that the identities in Eqs.~\eqref{Medellin} and \eqref{xtombasis} imply 
\bea
\int  dx_3  \, x_3^{h_3+m_3-1} \lim_{x_1\rightarrow \infty} x_1^{2(h_1-1)} && \braket{\of{j V_{h_1}}(x_1) V_{h_2}(x_2=0) V_{h_3}(x_3)}\label{253} \\[1ex]
&& \qquad \qquad  = \braket{\of{j V_{h_1}}_{h_1-1,h_1-1} V_{h_2,-h_2} V_{h_3,m_3}}, \nn
\eea
with 
\beq
\of{j V_h}_{h-1,m} = - j^+ V_{h,m-1} + 2j^0V_{h,m}- j^- V_{h,m+1}\label{jVm}.
\eeq
Using \eqref{251} on the first line of \eqref{253} it turns out that, up to the $z$-dependence,
\beq
C(h_i) \delta(h_1-1-h_2+m_3) = \braket{\of{j V_{h_1}}_{h_1-1,h_1-1} V_{h_2,-h_2} V_{h_3,m_3}}\label{255}
\eeq
We are then interested in $m$-basis correlators of the form 
\beq
\braket{\off{j^a V_{h_1,m_1}} V_{h_2,m_2} V_{h_3,m_3}}\label{2}.
\eeq
By using 
\beq
\off{j^a V_{h,m}}(z) = \oint_{z} dw \frac{1}{w-z} j^a(w) V_{h.m}(z), 
\eeq
and reversing the contour, the OPEs in \eqref{JV OPE} imply
\bea
&\braket{\off{j^a V_{h_1,m_1}} V_{h_2,m_2} V_{h_3,m_3}}&= \frac{(m_2-a(h_2-1))}{z_{12}}\braket{V_{h_1,m_1} V_{h_2,m_2+a} V_{h_3,m_3}}\\
&&+\frac{(m_3-a(h_3-1))}{z_{13}}\braket{V_{h_1,m_1} V_{h_2,m_2} V_{h_3,m_3+a}}. \nn
\eea
Consequently, given that $(jV_{h_1})_{h_1-1,h_1-1} =  - j^-V_{h_1,h_1}$,
\beq
\braket{\of{j V_{h_1}}_{h_1-1,h_1-1} V_{h_2,-h_2} V_{h_3,m_3}} = \frac{z_{23}}{z_{12}z_{13}}(h_1-h_2-h_3)\braket{V_{h_1,h_1} V_{h_2,-h_2} V_{h_3,m_3-1}},
\eeq
where we have used the relation
\beq
\braket{V_{h_1,h_1} V_{h_2,-h_2-1} V_{h_3,m_3}} = (m_3+h_3-1)\braket{V_{h_1,h_1} V_{h_2,-h_2} V_{h_3,m_3-1}}.
\eeq
Applying the same strategy we can re-derive the relevant $m$-basis three-point function \cite{Satoh:2001bi} as
\beq
\int x_3^{m_3-1+h_3-1}\lim_{x_1\rightarrow \infty} \lim_{ x_2\rightarrow 0}x_1^{2h_1}\braket{V_{h_1}(x_1)V_{h_2}(x_2)V_{h_3}(x_3)} = \braket{V_{h_1,h_1} V_{h_2,-h_2} V_{h_3,m_3}}.
\eeq
The $x$-basis three-point function is given by
\beq
\braket{V_{h_1}(x_1)V_{h_2}(x_2)V_{h_3}(x_3)} = C_H(h_i) x_{12}^{h_3-h_2-h_1}x_{23}^{h_1-h_2-h_3}x_{13}^{h_2-h_3-h_1}
\eeq
where, again, we have ignored the $z$-dependence. Therefore
\beq
\braket{V_{h_1,h_1} V_{h_2,-h_2} V_{h_3,m_3}} =  \frac{C_H(h_i)\delta(h_1-1-h_2-m_3)}{z_{12}^{\Delta_{h_1}+\Delta_{h_2}-\Delta_{h_3}}z_{23}^{\Delta_{h_2}+\Delta_{h_3}-\Delta_{h_1}}z_{13}^{\Delta_{h_1}+\Delta_{h_3}-\Delta_{h_2}}}.
\eeq
Note that the delta function coincides with that of \eqref{255}. Hence, we find $C(h_i) = (h_1-h_2-h_3)C_H(h_i)$. In other words, 
\beq
\braket{\of{j V_{h_1}}(x_1) V_{h_2}(x_2) V_{h_3}(x_3)} = (h_1-h_2-h_3)\frac{x_{12}x_{13}}{x_{23}}\frac{z_{23}}{z_{12}z_{13}}\braket{V_{h_1}(x_1) V_{h_2}(x_2) V_{h_3}(x_3)}\label{4},
\eeq
which agrees with the results of \cite{Dabholkar:2007ey}. 
This result can also be used to compute the fermionic correlator with an extra $\hat{j}(x)$ insertion. Setting $h_1 = 0$ (for the identity operator) and $h_2=h_3=-1$ in Eq.~\eqref{4}, we get  
\beq
\braket{\hat{j}(x_1)\psi(x_2)\psi(x_3)} = 2 \, \frac{x_{12}x_{13}}{x_{23}}\frac{z_{23}}{z_{12}z_{13}}\braket{\psi(x_2)\psi(x_3)}.
\eeq
Analogous expressions hold for the SU(2) correlators. 

We have now obtained all the necessary ingredients of the (unflowed) supersymmetric three-point function, which gives 
\bea
& &\braket{\cV^{(0)}_{h_1}(x_1,y_1) \cV_{h_2}  (x_2,y_2)\cV_{h_3} (x_3,y_3)}  =   \\[1ex] 
& & \qquad  n_5\of{2-h} \frac{x_{12}x_{23}x_{13}}{z_{12}z_{23}z_{13}}  \braket{V_{h_1}(x_1)V_{h_2}(x_2)V_{h_3}(x_3)}\braket{W_{h_1-1}(y_1)W_{h_2-1}(y_2)W_{h_3-1}(y_3)}, \nn
\eea
where we have defined $h = h_1+h_2+h_3$. 
As it was discussed in \cite{Dabholkar:2007ey,Gaberdiel:2007vu}, the shifts in the bosonic levels of SL(2,$\R$)$_{n_5+2}$ and SU(2)$_{n_5-2}$ conspire precisely so that, for operators in the discrete representations with $h_i = l_i + 1$, the product of the corresponding three-point functions  simplify considerably. Explicitly, and omitting the $x$ and $y$ dependence,
\beq
\braket{V_{h_1}V_{h_2}V_{h_3}}\braket{W_{h_1-1}W_{h_2-1}W_{h_3-1}} \equiv D(h_i) = \frac{\nu^{1-h}}{2\pi^2b^4\sqrt{\gamma(b^2)}}\prod_{i=1}^{3}\frac{1}{\sqrt{\gamma(b^2(2h_i-1))}}.
\label{SL2xSU2}
\eeq
Consequently, one gets 
\bea
& &\braket{\cV^{(0)}_{h_1}(x_1,y_1)\cV_{h_2}(x_2,y_2)\cV_{h_3}(x_3,y_3)} =  \\[1ex] 
& & \qquad  \frac{n_5\of{2-h}D(h_i)}{z_{12}z_{23}z_{13}}\of{\frac{y_{12}}{x_{12}}}^{H_1+H_2-H_3}\of{\frac{y_{23}}{x_{23}}}^{H_2+H_3-H_1}\of{\frac{y_{13}}{x_{13}}}^{H_1+H_3-H_2}, \nn
\eea
where, as before, $H_i=h_i-1$. One can compute all other NSNS correlators involving $\cV$ and $\cW$ insertions analogously, see \cite{Dabholkar:2007ey,Gaberdiel:2007vu}.


In order to identify the precise boundary duals one has to integrate over the worldsheet insertion points. It was shown in \cite{Maldacena:2001km,Dabholkar:2007ey} that this procedure removes the divergence appearing in the two-point functions \eqref{2ptVV} and \eqref{2ptWW} coming from the $\delta(h_1-h_2)$ factors, but it also generates a finite multiplicative factor $1-2h$. This, together with the correct normalization for the spacetime operators, can be derived by making use of the spacetime Ward identities. We review this computation in Sec.~\ref{sec:normalization} below, where we further provide the extension to the spectrally flowed sectors.

Finally, in order to compare the three-point worldsheet three-point functions with those of the holographic CFT we also need to include the usual factors associated with the string coupling $g_s$ and the volume of the $T^4$. 
As discussed in \cite{Dabholkar:2007ey,Gaberdiel:2007vu}, this leads to a precise matching with the correlators of the symmetric product orbifold CFT. At this point of the moduli space, the operators defined above correspond to families of chiral primaries of twist $n=2h-1$ and weights $H=(n\pm 1)/2$, respectively. 

However, due to Eq.~\eqref{rangeSL2R}, operators in the unflowed sector of the worldsheet theory only account for chiral primaries with twists $n < n_5$. Except for the rather special cases where $n \in n_5 \mathbb{N}$ \cite{Seiberg:1999xz} (see also \cite{Eberhardt:2018vho}), all primaries with higher twists live in the spectrally flowed sectors of the worldsheet theory \cite{Giribet:2007wp}, which now turn to.

\section{Spectrally flowed states}

We now discuss states belonging to the spectrally flowed representations in the NSNS sector of the worldsheet theory. These include both short strings, which encompass most of the spacetime chiral primaries, and also long strings which remain at finite energy when reaching the asymptotic boundary of AdS$_3$, and for which the spectral flow charge is understood as a winding number.

\subsection{Vertex operators for short strings}

When considering the discrete representations it is useful to perform spectral flow not only in the SL(2,$\R$) sector but also in SU(2) \cite{Giribet:2007wp}. 
For the latter, the spectral flow maps different standard affine representations into each other, while for SL(2,$\RR$) it generates new inequivalent representations, for which 
the conformal weight is not bounded from below \cite{Maldacena:2000hw}. 

The relevant spectral flow isomorphisms act on the current modes as
\beq
J^{0,\omega}_{n} = J^0_n - \frac{n_5}{2}\omega \delta_{n,0} 
\,, \quad
K^{0,\omega}_{n} = K^0_n + \frac{n_5}{2}\omega \delta_{n,0} 
\,, \quad
J^{\pm,\omega}_{n} = J^\pm_{n\pm\omega}
\,, \quad
K^{\pm,\omega}_{n} = K^\pm_{n\pm\omega}\, ,
\eeq 
while the fermions transform as
\beq
\psi^{0,\w}_n =\psi^0_n
\,, \qquad 
\chi^{0,\w}_n =\chi^0_n 
\,, \qquad \psi^{\pm,\omega}_n = \psi^\pm_{n\pm\omega}
\,, \qquad 
\chi^{\pm,\omega}_n = \chi^\pm_{n\pm\omega}.
\eeq
with $\omega \in \ZZ$. The currents $j^a, k^a,\hat{j}^a$ and $\hat{k}^a$ transform similarly with $n_5$ replaced by the corresponding levels.  The Virasoro modes shift according to 
\begin{subequations}
\bea
&L^{AdS_3}_{n} &= L^{AdS_3,\omega}_n - \omega J^{0,\omega}_n- \frac{n_5\omega^2}{4}\delta_{n,0},\\
&L^{S^3}_{n} &= L^{S^3,\omega}_n - \omega K^{0,\omega}_n + \frac{n_5\omega^2}{4}\delta_{n,0},
\eea
\label{ViraoromodesW}
\end{subequations}
while for the (total) matter supercurrent we have 
\beq 
G_r = G^{\omega}_r - \omega (\psi^{0,\omega}_r + \chi^{0,\omega}_r)\label{Gw}.
\eeq

Let us focus on the SL(2,$\R$) sector. Flowed vertex operators in the $m$-basis are constructed upon the spectrally flowed primaries, whose bosonic part we denote by  $V^{\omega}_{hm}$. These are primary fields with respect to the flowed currents $j^{a,\omega}$. Hence, for $\w>0$ they satisfy the following OPEs:
\begin{subequations}
\bea
&j^{0}(z)V^{\omega}_{h,m}(w)&\sim (m+\frac{k}{2}\omega)\frac{V^{\omega}_{h,m}(w)}{z-w}\, ,\\[1ex]
&j^{-}(z)V^{\omega}_{h,m}(w)&\sim 0 \, ,\label{J-Vw OPE}\\
&j^{+}(z)V^{\omega}_{h,m}(w)&\sim (m-h+1)\frac{V^{\omega}_{h,m+1}(w)}{(z-w)^{1+\omega}}+\sum_{l=0}^{\omega-1}\frac{\off{j^{0}_l V^{\omega}_{h,m}}(w)}{(z-w)^{l+1}} \, . \label{J+Vw OPE}
\eea
\label{JVwOPE}
\end{subequations}
Moreover, $V^{\omega}_{h,m}$ has conformal weight 
\beq
\Delta^\omega_h = -\frac{h(h-1)}{n_5} - \omega m - \frac{k \omega^2}{4}, 
\label{deltahW}
\eeq
and it is the lowest-weight state in a representation of the zero-mode algebra with spin
\begin{equation}
    h_\omega = m + k\omega/2.
    \label{hWdef}
\end{equation} 
Similarly, $V^{-\omega}_{h,m}$ corresponds to a highest-weight state with spin $h_\omega = -m + k \omega/2$. Note also that $V^{\omega}_{h,m}$ and $V^{-\omega}_{h,-m}$ (with $\w>0$) have the same spin; they will contribute to the same $x$-basis operator.

Flowed fermionic fields conveniently expressed in terms of their bosonized form as  
\beq
\psi^{-,\omega} = \sqrt{n_5}e^{-i(1+ \omega)H_1 }\qqquad\psi^{+,\omega} =\sqrt{n_5}e^{i(1- \omega)H_1 }\qqquad\psi^{0,\omega} = \psi^0 e^{-i\omega H_1}.
\label{flowedfermionsSL2}
\eeq
These correspond to lowest-weight states in spin $a-\omega$ representations, respectively, with $a=0,+,-$. Note that, for any $\omega>0$, the flowed fermions  have either negative or zero spin, and  belong to a finite representation of the global part of the symmetry algebra. Also, 
$\psi^{0,\omega=1} = \frac{\sqrt{n_5}}{2}\hat{j}^-$, hence $\psi^{0,\omega+1} = \frac{\sqrt{n_5}}{2}\hat{j}^{-,\omega}$. Moreover, the fermionic identity is mapped to a field of spin $-\omega$ and conformal weight $\omega^2/2$ given by
\beq
\mathds{1}^{\omega} = \frac{1}{\sqrt{n_5}}\psi^{-,\omega-1} = e^{-i\omega H_1}.
\eeq 

Combining the above expressions, we see that  flowed primary states of the complete supersymmetric SL(2,$\RR$)$_{n_5}$ algebra are of the form 
\beq
\of{\psi V_h}^\omega_{h-1,m} = - \psi^{+,\omega} V^{\omega}_{h,m-1} + \psi^{0,\omega} V^{\omega}_{h,m}-\psi^{-,\omega} V^{\omega}_{h,m+1}.
\eeq
By including the analogous SU(2) factors, this leads to vertex operators given by
\beq
 e^{-\varphi} \of{\psi V_h}^\omega_{h-1,m} W^{\omega}_{l,n} e^{-i\omega H_2}.
\eeq
However, not all of these are physical. From Eqs.~\eqref{ViraoromodesW} and \eqref{Gw} one finds that BRST-invariant operators must satisfy $l = h-1$ and $m = -n = \pm (h-1)$, thus giving \cite{Giribet:2007wp} 
\begin{subequations}
\bea
&\cV^{\omega}_{h,\pm h} &= e^{-\varphi}\psi^{\mp,\omega}V^\omega_{h,\pm h} W^{\omega}_{h-1,\mp (h-1)} e^{-i\omega H_2} \, , \\
&\cW^{\omega}_{h,\pm h} &= e^{-\varphi}e^{-i\omega H_1}V^{\omega}_{h,\pm h}W^{\omega}_{h-1,\mp (h-1)}\chi^{\mp,\omega}\,,
\eea
\label{Vwm}
\end{subequations}
where we have included the analogous construction based upon the unflowed states polarized in the SU(2) directions.

The corresponding $x$-basis operators are constructed analogously to those of the  unflowed sector, that is, by constructing the appropriate linear combination of all fields in the associated global multiplet. These are obtained by acting freely with $J_0^\pm$ on the highest/lowest-weight states defined above. In the bosonic SL(2,$\R$) sector, we have 
\beq
V^\omega_{h}(x) = \sum_{\tilde{m}=h+k\omega/2+n}^\infty x^{-(h+k\omega/2) - \tilde{m}}
\Uu^{\omega}_{h,\tilde{m}} \qqquad 
\Uu^{\omega}_{h,\pm (h+k\omega/2)} = V_{h, \pm h}^{\pm\w}.
\eeq
Note, however, that for $\w \neq 0$, the modes $\Uu^{\omega}_{h, \tilde{m}}$ with $\tilde{m} \neq \pm (h+k\omega/2)$ have no simple $m$-basis expressions.
As anticipated above, the $x$-basis field contains both the lowest- and highest-weights representations, i.e.~both those with positive and negative spectral flow charges.
It follows from Eqs.~\eqref{JVwOPE} that the zero-mode currents still act on the flowed $x$-basis operators through the differential operators  $D^a_{x,h+k\omega/2}$ defined in \eqref{zero modes}.  
An analogous construction can be done for the fermions and the fermionic currents: 
\begin{subequations}
\label{Vw+-}
\bea
&&\psi^{\omega}(x) = \sum_{\tilde{m}=-(1+\omega)}^{1+\omega} x^{(1+\omega)-\tilde{m}}\hat{\Uu}^{\omega}_{-1,\tilde{m}},\quad  \hat{\Uu}^{\omega}_{-1,\mp(1+\omega)} = \psi^{\mp,\pm \omega} = \sqrt{n_5}e^{\mp i(\omega+1)H_1},\\[1ex]
&&\hat{j}^{\omega}(x) = \frac{2}{\sqrt{n_5}} \sum_{\tilde{m}=-(1+\omega)}^{1+\omega} x^{(1+\omega)-\tilde{m}}\hat{\Uu}^{0,\omega}_{\hat{h}_\omega,\tilde{m}},\quad  \hat{\Uu}^{0,\omega}_{-1,\mp (1+\omega)} = \psi^{0} e^{\mp i(\omega+1)H_1}.
\eea
\end{subequations}

The supersymmetric $x$-basis vertex operators come in four families, depending on the sign choice in Eq.~\eqref{Vwm} and also on the polarization. On the one hand, we have
\begin{subequations}
\bea
&&\cV^{\omega}_{h}(x,y) = \frac{1}{\sqrt{n_5}}e^{-\varphi} \psi^{\omega}(x) V^{\omega}_{h}(x) W^{\omega}_{h-1}(y) \chi^{\omega-1}(y)\\
&&\cV^{\omega}_{-h}(x,y) = \frac{1}{\sqrt{n_5}}e^{-\varphi} \psi^{\omega-2}(x) V^{\omega}_{-h}(x) W^{\omega}_{-(h-1)}(y) \chi^{\omega-1}(y),
\eea\label{5}
\end{subequations}
where we defined $W_l(y)$ and  $\chi^{\omega}(y)$ analogously to their SL(2,$\R$) counterparts, and used  $\psi^{+,\omega}= \psi^{-,\omega-2}$. 
Here $V^{\omega}_{\pm h}(x)$ are the $x$-basis bosonic primary fields of spin $h_\omega = \pm h + k\omega/2$, defined in such a way that the corresponding lowest- and highest-weight modes can be extracted of the vertex  as follows:
\begin{subequations}
\bea
\lim_{x\rightarrow \infty}x^{2(k\omega/2 +h)} V^{\omega}_h(x) = V^{\omega}_{h,h} \qqquad &&\lim_{x\rightarrow 0} V^{\omega}_h(x) = V^{-\omega}_{h,-h}\,,\\
\lim_{x\rightarrow \infty}x^{2(k\omega/2 -h)} V^{\omega}_{-h}(x) = V^{\omega}_{h,-h} \qqquad &&\lim_{x\rightarrow 0} V^{\omega}_{-h}(x) = V^{-\omega}_{h,h}\, .
\eea
\label{limitsVwx}
\end{subequations}
Similarly, we also have 
\begin{subequations}
\label{Ww+-}
\bea
&&\cW^{\omega}_{h}(x,y) = \frac{1}{\sqrt{n_5}}e^{-\varphi} \psi^{\omega-1}(x) V^{\omega}_{h}(x) W^{\omega}_{h-1}(y) \chi^{\omega}(y),\\
&&\cW^{\omega}_{-h}(x,y) = \frac{1}{\sqrt{n_5}}e^{-\varphi} \psi^{\omega-1}(x) V^{\omega}_{-h}(x) W^{\omega}_{-(h-1)}(y) \chi^{\omega-2}(y).
\eea
\end{subequations}
However, and as we discuss next, only two of these four families of vertex operators are actually independent.


\subsection{Series Identifications}
\label{Sec:SeriesIdenfifications}

As is well known, for short strings the independence of the spectrally flowed representations holds only up to the identifications \cite{Maldacena:2001km}
\beq
V^{\omega}_{hh} = R(h)V^{\omega+1}_{\frac{k}{2}-h,-\of{\frac{k}{2}-h}} \qqquad W^{\omega}_{l,-l} = W^{\omega+1}_{\frac{k'}{2}-l,\frac{k'}{2}-l}
\eeq
where the proportionality factor for the SL(2,$\R$) case takes the form  
\beq
R(h) = \pi b^2 \nu^{k/2} B(h),
\eeq
as can be derived from the two-point function, using that
\beq
B(h) B\left(\frac{k}{2}-h\right) = \of{\pi b^2 \nu^{k/2}}^{-2}.
\eeq
In other words, $R(h) \sim B(h) \sim B\left(\frac{k}{2}-h\right)^{-1}$ up to $h$-independent but $k$-dependent factors. 
In the supersymmetric theory, this translates into \cite{Martinec:2020gkv}
\begin{equation}
    \cV^{\omega}_{h,h}= R(h)\cW^{\omega+1}_{\frac{k}{2}-h,-\of{\frac{k}{2}-h}}
    \qqquad
    \cW^{\omega}_{h,h}= R(h)\cV^{\omega+1}_{\frac{k}{2}-h,-\of{\frac{k}{2}-h}}.
    \label{SusySeriesIdentification}
\end{equation}

As it was shown recently in \cite{Iguri:2022eat} for the bosonic SL(2,$\R$) case, these identifications also hold in for $x$-basis operators involving the global descendants. This can be understood intuitively as follows. Vertex operators in the $x$-basis can be thought of as a \textit{translated} version of the $m$-basis flowed primaries associated to their values at the origin $x=0$, see Eqs.~\eqref{limitsVwx} \cite{Eberhardt:2019ywk}. Indeed, translations in the complex $x$-plane are generated by the action of $J_0^- \sim \der_x$. A similar discussion holds for the SU(2) primaries and the respective fermions. As a consequence, we find that the supersymmetric vertex operators satisfy the following identities:
\beq
\cV^{\omega}_{h}(x,y) = R(h) \cW^{\omega+1}_{-\of{\frac{n_5}{2}-h+1}}(x,y) \qqquad \cW^{\omega}_{h}(x,y) = R(h) \cV^{\omega+1}_{-\of{\frac{n_5}{2}-h+1}}(x,y).
\label{SusyIdentif}
\eeq
Hence, we can focus our attention on the computation of correlators involving only  $\cV^{\omega}_{h}$ and/or  $\cW^{\omega}_h$ insertions, the remaining ones being fixed by means of the series identifications in Eq.~\eqref{SusyIdentif}. 
\subsection{Two-point functions}
With the definitions given in the previous section, it is possible to compute the two-point functions straightforwardly as in the unflowed sector. The bosonic two-point functions can be computed by taking the limits
\beq
\lim_{x_1\rightarrow 0}\lim_{x_2\rightarrow \infty}x_1^{2h_{\omega_1}}\braket{V^{\omega_1}_{h_2}(x_1)V^{\omega_2}_{h_2}(x_2)} = \braket{V^{\omega_1}_{h_1,h_1}V^{-\omega_2}_{h_2,-h_2}}
\eeq
where $h_{\omega_i} = h_i + k\omega_i/2$. The $m$-basis two-point function must conserve the total amount of spectral flow,  $\omega_1 = \omega_2$ \cite{Maldacena:2001km}. Additionally, and as can be shown by using the parafermionic decomposition, the $m$-basis correlators only depend on the total spectral flow charge, but not on the specific assignment of spectral flow charges. This is up to the shifts in the usual powers of $z_{ij}$. Then
\beq
\braket{V^{\omega_1}_{h_1}(x_1)V^{\omega_2}_{h_2}(x_2)} = \frac{\delta(h_1-h_2)\delta_{\omega_1,\omega_2}}{z_{12}^{2\Delta^{\omega_1}_{h_1}}x_{12}^{2h_{\omega_1}}}B(h_1),
\eeq
with $\Delta^\omega_h$ defined as in Eq. \eqref{deltahW}. Similarly, this can be done for the other sectors and we have
\bea
&& \braket{\cV^\omega_{h_1}(x_1)\cV^\omega_{h_2}(x_2)}  = n_5\frac{\delta(h_1-h_2)}{z_{12}^{2}}\of{\frac{y_{12}}{x_{12}}}^{2H_{\omega_1}}B(h_1),\quad H_{\omega_1} = h_1-1+\frac{n_5}{2}\omega,\\
&&  \braket{\cW^\omega_{h_1}(x_1)\cW^\omega_{h_2}(x_2)} = n_5\frac{\delta(h_1-h_2)}{z_{12}^{2}}\of{\frac{y_{12}}{x_{12}}}^{2H_{\omega_1}}B(h_1), \quad H_{\omega_1} = h + \frac{n_5}{2}\omega,
\eea
where we have used that
\beq
\braket{W^\omega_{l}(y_1)W^\omega_{l}(y_2)}= \frac{y_{12}^{2l+k'\omega}}{z_{12}^{2l(l+1)/n_5 + 2l\omega + k'\omega^2/2}},
\eeq
and 
\beq 
\braket{\psi^\omega(x_1)\psi^\omega(x_2)} = n_5 \frac{x_{12}^{2(\omega+1)}}{z_{12}^{(\omega_1+1)^2}}\qqquad
\braket{\chi^\omega(x_1)\chi^\omega(x_2)} = n_5 \frac{y_{12}^{2(\omega+1)}}{z_{12}^{(\omega_1+1)^2}}.
\eeq
The correlators involving the vertex operators $\cV^\omega_{-h}$ and $\cW^\omega_{-h}$ can be derived by the series identification given in Eq. \eqref{SusyIdentif}.

\subsection{Picture Changing}

As in the unflowed case, in order to compute string three-point we will need the picture-zero version of the spectrally flowed vertex operators \cite{Giribet:2007wp}. We first discuss the flowed primaries and then consider the $x$-basis operators. 

The SL(2,$\RR$) relevant contributions to the supercurrent $G$ can be expressed in terms of the $H_I$ bosons as 
\begin{subequations}
\bea
&G_{AdS_3} &= \frac{2}{n_5}\off{ \frac{\sqrt{n_5}}{2}\of{e^{iH_1}j^- + e^{-iH_1}j^+}-\psi^0j^0-i\psi^0\p H_1} \, , \\
&G_{S^3} &= \frac{2}{n_5}\off{ \frac{\sqrt{n_5}}{2}\of{e^{iH_2}k^- + e^{-iH_2}k^+}+\chi^0k^0 +i\chi^0\p H_2} \, .
\eea
\end{subequations}
Using Eqs.~\eqref{PictureChangingOp} and \eqref{Gw}, one finds that 
\beq
\cV^{\omega,(0)}_{h,h} = \mathcal{A}^{-,\omega}_1 + (-1)^{\omega+1}\mathcal{A}^{-,\omega}_2 \, ,
\eeq
with 
\begin{subequations}
\bea
&\mathcal{A}^{-,\omega}_1 &=\off{\frac{1}{n_5}j^{-,\omega}\psi^{-,\omega-1} - \frac{1}{\sqrt{n_5}}H_\omega \hat{j}^{-,\omega}}V^{\omega}_{h,h}W^{\omega}_{h-1,-(h-1)}\chi^{-,\omega-1}\, ,\\
&\mathcal{A}^{-,\omega}_2 &= \frac{1}{\sqrt{n_5}}V^{\omega}_{h,h}\left[ W^{\omega}_{h-1,-(h-2)}\psi^{-,\omega}\chi^{-,\omega} + 
H_\omega W^{\omega}_{h-1,-(h-1)} \psi^{-,\omega}\hat{k}^{-,\omega-1}\right],
\eea
\end{subequations}
where $H_\omega = h - 1 + n_5\omega/2$ total spacetime weight. 
In the $x$-basis, this reads 
\beq
\cV^{\omega,(0)}_{h}(x,y) = \mathcal{A}^{\omega}_1(x,y) + (-1)^{\omega+1}\mathcal{A}^{\omega}_2(x,y)
\eeq
with
\begin{subequations}
\bea
&\mathcal{A}^{\omega}_1 (x,y) &=\frac{1}{n_5}\off{j^{\omega}(x)
\psi^{\omega-1}(x) - \sqrt{n_5}H_\omega \hat{j}^{\omega}(x)}V^{\omega}_{h}(x)W^{\omega}_{h-1}(y)\chi^{\omega-1}(y)
\,,\\
&\mathcal{A}^{\omega}_2(x,y) &= \frac{\psi^{\omega}(x)}{\sqrt{n_5}}V^{\omega}_{h}(x) \left[  W^{\omega}_{h-1,-(h-2)}(y)\chi^{\omega}(y) + H_\omega W^{\omega}_{h-1}(y) \hat{k}^{\omega-1}(y) \right]\,,
\eea
\end{subequations}
where 
\beq
j^{\omega}_n(x) = j^{+}_{n-\omega} - 2x j^0_{n-\omega} + x^2 j^{-}_{n-\omega} \, .
\label{6}
\eeq
Here $W^{\omega}_{h-1,-(h-2)}(y)$ are the sum over the multiplet constructed from the flowed primary $W^{\omega}_{h-1,-(h-2)}$. Note that, both in the $\psi$ and $\chi$ sectors, the fermion number on each factor of $\cA_1$ differs in one unit with respect to those in $\cA_2$. Hence, whenever $\cA^\omega_1$ leads to a non-zero contribution in a given correlator, that of $\cA^{\omega}_2$ will vanish, and vice-versa. Finally, an analogous discussion gives 
\begin{equation}
    \cW^{\omega,(0)}_{h} = \mathcal{B}^{\omega}_1(x,y) + (-1)^{\omega}\mathcal{B}^{\omega}_2(x,y),
    \label{WwPicZero}
\end{equation}
where
\begin{subequations}
\bea
&\mathcal{B}^{\omega}_1(x,y) &= \frac{1}{\sqrt{n_5}}\left[  V^{\omega}_{h,h+1}(x)\psi^{\omega}(x) -H_{\omega}V^{\omega}_{h}(x) \hat{j}^{\omega-1}(x) \right]\chi^{\omega}(y)W^{\omega}_{h-1}(y) 
\,,\\
&\mathcal{B}^{\omega}_2 (x,y) &=\frac{1}{\sqrt{n_5}}\psi^{\omega-1}(x)\off{\chi^{\omega-1}(y)k^{\omega}(y)
 - H_\omega \hat{k}^{\omega}(x)}V^{\omega}_{h}(x)W^{\omega}_{h-1}(y)\,, 
\eea
\end{subequations}
with
\beq
k^{\omega}(y)_n = k^+_{n-\omega}-2y k^0_{n-\omega}-y^2k^-_{n-\omega}.
\eeq

It follows from the above expressions that generic NS sector three-point functions will involve bosonic correlator of the form 
\beq
\braket{\off{j^{\omega_1}V^{\omega_1}_{h_1}}(x_1)V^{\omega_2}_{h_2}(x_2)V^{\omega_3}_{h_3}(x_3)} 
\qqquad 
\braket{\off{k^{\omega_1}W^{\omega_1}_{h_1-1}}(y_1)W^{\omega_2}_{h_2-1}(y_2)W^{\omega_3}_{h_3-1}(y_3)}.
\eeq
To the best of our knowledge, these have not been computed in the literature. One of the main results of this paper is to obtain closed-form expressions for all such $m$-basis correlators. We will also compute several families of their $x$-basis counterparts. These computations are performed in section \ref{sec: flowed 3point}. 

\subsection{Vertex operators for long strings}
The long string sector of the model is built upon vertex operators belonging to the continuous representation of the SL(2,$\RR$)$_{k}$ algebra. The simplest vertex operator that can be constructed by this is the non-excited state given by
\beq
\cZ_{h,m;l,n} = e^{-\varphi}V_{h,m}W_{l,n}.
\eeq
This field does not satisfy the GSO projection. However, we will consider the flowed versions  
\beq
\cZ^\omega_{h,m;l,n} = \frac{1}{\sqrt{n_5}}e^{-\varphi}\psi^{-,\omega-1}V^\omega_{h,m}W_{l,n}.\label{Zw}
\eeq
For odd $\omega$ this is a physical field allowed by the GSO projection. (A similar construction can be carried out for even spectral flow charges.) The Virasoro condition reads
\beq
-\frac{h(h-1)}{n_5}-\omega m -\frac{n_5\omega^2}{4} + \frac{l(l+1)}{n_5} = \frac{1}{2}.
\eeq
For states in the continuous representation, the SL(2,$\RR$)$_k$ projection $m$ can take any real value. Therefore, the Virasoro condition can be satisfied for all values of $h$, $l$ and $\omega$. Note that, unlike the short-string states of the previous section, long strings do not require a relation between the SL(2,$\RR$)$_k$ and SU(2)$_{k'}$ spin. The vertex operator given by Eq. \eqref{Zw} has spacetime weight $H_\omega = m + n_5 \omega /2$ ($\omega > 0$). The $x$-basis field is  built as 
\beq
\cZ^\omega_{h,m;l}(x,y) = \frac{1}{\sqrt{n_5}}e^{-\varphi}\psi^{\omega-1}(x)V^\omega_{h,m}(x)W_l(y).
\eeq
Note that we must keep track of the value of $m$ even in the $x$-basis since the spacetime weight depends directly on it. As before, the $x$-basis field contains both positive and negative flow representations and satisfies that
\bea
&&\lim_{x\rightarrow \infty} x^{2H_\omega} \cZ^\omega_{h,m;l}(x,y) = \frac{1}{\sqrt{n_5}}e^{-\varphi}\psi^{-,\omega-1}V^\omega_{h,m}W_l(y),\\
&&\lim_{x\rightarrow 0} \cZ^\omega_{h,m;l}(x,y) = \frac{1}{\sqrt{n_5}}e^{-\varphi}\psi^{+,-(\omega-1)}V^{-\omega}_{h,-m}W_l(y).
\eea
The picture zero version of the long strings operator is given by
\beq
\cZ^{\omega,(0)}_{h,m;l,n} = \eta_{ab}\psi^{a,\omega}g^{b}_{h,m}V^{\omega}_{h,m+b}W_{l,n} + \delta_{\alpha\beta}\psi^{-,\omega-1}\chi^{\alpha}g^{\beta}_{l,n}V^{\omega}_{h,m}W_{l,n+\beta},\label{Zwzero}
\eeq
where we assume an implicit sum over $a,b$ and $\alpha,\beta$, and with 
\bea
&&g^{b}_{hm} = \of{m+h-1,m+\frac{n_5}{2}\omega,m-h+1},\\
&&g^{\beta}_{ln} = \of{l+1-n,n,l+1+n}, 
\eea
where $b = -,0,+$. Note that both terms in \eqref{Zwzero} have the same total fermion number $\omega +1 $,  but they differ in one unit with respect to fermion numbers in the SL(2,$\RR$) and SU(2) sectors. Consequently, similarly to the short strings picture zero operators of the previous section, both terms are mutually exclusive inside a correlator.

\section{Correlators, currents and spectral flow}
\label{sec: flowed 3point}

As discussed above, the computation of the three-point functions in the NSNS sector of the worldsheet theory involves bosonic correlation functions with descendant insertions. These are non-trivial in the presence of spectrally flowed vertex operators. Due to the complicated OPEs between the latter and the currents, these correlators cannot be obtained by the standard contour integration methods. In this section, which contains our main results, we present a strategy to compute some families of such correlators. 

\subsection{Bosonic correlators with current insertions and general $m$-basis results}\label{sec: Bosonic Currents}

Let us focus on the SL(2,$\R$) case. Our goal is to compute 
\beq
\braket{\of{j^{\omega_1}V^{\omega_1}_{h_1}}(x_1)V^{\omega_2}_{h_2}(x_2)V^{\omega_3}_{h_3}(x_3)}\label{current correlator}.
\eeq
In order to do so, we will follow a similar strategy to that used in the unflowed case, see Sec.~\ref{sec: unflowed 3point}. The $x$- and $z$-dependence of \eqref{current correlator} is fixed by the global Ward identities. Indeed, the zero-mode currents act on the operator $\of{j^\omega V^\omega_{h}}$ in terms of its worldsheet conformal dimension and spin, namely  
\begin{equation}
 \tilde{\Delta}^\omega_h = \Delta^{\omega}_h + \omega + 1.
 \qqquad \tilde{h}_\omega =   h_\omega - 1\, ,
\end{equation}
where $\Delta_h^\omega$ and $h_\omega$ were defined in Eqs.~\eqref{deltahW} and \eqref{hWdef} with $m=h$, respectively, and we have used \eqref{6}. 
Hence, we can write 
\beq
\braket{\off{j^{\omega_1}V^{\omega_1}_{h_1}}(x_1)V^{\omega_2}_{h_2}(x_2)V^{\omega_3}_{h_3}(x_3)} = C^{\omega_i}(h_i)\frac{x_{12}^{h_{\omega_3}-\tilde{h}_{\omega_1}-h_{\omega_2}}x_{23}^{\tilde{h}_{\omega_1}-h_{\omega_2}-h_{\omega_3}}x_{13}^{h_{\omega_2}-h_{\omega_3}-\tilde{h}_{\omega_1}}}{z_{12}^{\tilde{\Delta}^{\omega_1}_{h_1}+\Delta^{\omega_2}_{h_2}-\Delta^{\omega_3}_{h_3}}z_{23}^{\Delta^{\omega_3}_{h_3}+\Delta^{\omega_2}_{h_2}-\tilde{\Delta}^{\omega_1}_{h_1}}z_{13}^{\tilde{\Delta}^{\omega_1}_{h_1}+\Delta^{\omega_3}_{h_3}-\Delta^{\omega_2}_{h_2}}}\, ,
\label{corr1}
\eeq
such that we only need to determine the structure constants $C^{\omega_i}(h_i)$.

As in the unflowed case, the goal is to compute these constants by  relating the correlator in Eq.~\eqref{corr1} with some $m$-basis three-point functions.
Generically, we are interested in correlators of the form  
\beq
\braket{\off{j^{a,\omega_1} V^{\omega_1}_{h_1,m_1}}V^{\omega_2}_{h_2,m_2}V^{\omega_3}_{h_3,m_3}},
\eeq
where, from now on, $\omega_i$ is allowed to be positive or negative. These have not been computed in the literature. 

When the current involved is $j^0$, one can proceed analogously to the unflowed case. By writing the normal-ordered operator  $\off{j^{a,\omega_1} V^{\omega_1}_{h_1,m_1}}$ as a Cauchy integral and inverting the contour, we obtain 
\beq
\braket{\off{j^{0,\omega_1} V_{h_1,m_1}^{\omega_1}} V^{\omega_2}_{h_2,m_2} V^{\omega_3}_{h_3,m_3}}= \off{\frac{m_2+k\omega_2/2}{z_{12}}+\frac{m_3+k\omega_3/2}{z_{13}}}\braket{V^{\omega_1}_{h_1,m_1} V^{\omega_2}_{h_2,m_2}V^{\omega_3}_{h_3,m_3}}\,,
    \label{corr1m}
\eeq
showcasing the expected eigenvalues as shifted by spectral flow.  On the other hand, for the currents $j^\pm$ case one needs to take into account the pole structure appearing in the OPEs of Eqs.~\eqref{JVwOPE}.  $V^{\omega_i}_{h_i,m_i}$ and the currents $j^{a,\omega_1}$, that are non-trivial whenever $a= \pm$ and $\omega_i \neq \omega_1$. In these cases, the normal-ordered products appearing in \eqref{corr1m} are given by 
\beq
\off{j^{\pm,\omega_1}V^{\omega_1}_{h_1,m_1}}(z_1) = \oint_{z_1} dw \frac{j^{\pm,\omega_1}(w)}{(w-z_1)} V^{\omega_1}_{h_1,m_1}(z_1) = \oint_{z_1} dw \frac{j^{\pm}(w)}{(w-z_1)^{\mp\omega_1+1}} V^{\omega_1}_{h_1,m_1}(z_1).
\label{genJ}
\eeq
In order to compute \eqref{corr1m} for $a=\pm$, and for a given correlator, we define 
\beq
\cJ^{\pm,\omega_i}(z) \equiv j^{\pm}(z)\prod_i (z-z_i)^{\pm \omega_i}. \label{Jw}
\eeq
The powers of $(z-z_i)^{\pm \w_i}$ ensure that the modified currents $\cJ^{\pm,\omega_i}(z)$ behave near each flowed primary insertion analogously to unflowed currents near unflowed vertex operators. In particular, using the OPEs \eqref{J-Vw OPE} and \eqref{J+Vw OPE}, we find that
\beq
\oint_{\cC} \frac{dz}{z-z_1}\langle \cJ^{\pm,\omega_i}(z)V^{\omega_1}_{h_1,m_1}V^{\omega_2}_{h_2,m_2} V^{\omega_3}_{h_3,m_3} \rangle = 0  \label{7}, 
\eeq
when the contour $\cC$ encircles all three insertion points. To be precise, this holds only when there is no pole at infinity, but this happens for any configuration which satisfies the so-called $m$-basis spectral flow violation rules, i.e. whenever 
\begin{equation}
    |\w_1 + \w_2 + \w_3|\leq 1.
\end{equation}
On the other hand, we have 
\begin{align} 
&\oint_{z_1} dz \frac{\cJ^{\pm,\omega_i}(z)V^{\omega_1}_{h_1,m_1}(z_1)}{z-z_1} = \nn \\
&\qquad z_{12}^{\pm\omega_2}z_{13}^{\pm\omega_3}\offf{\off{j^{\pm,\omega_1}V^{\omega_1}_{h_1,m_1}}(z_1) \pm \of{\frac{\omega_2}{z_{12}}+\frac{\omega_3}{z_{13}}}(m_1\mp(h_1-1))V^{\omega_1}_{h_1,m_1\pm1}}.    
\end{align}
Note that we have picked up a contribution from the first two non-trivial terms in the corresponding OPEs \eqref{JVwOPE} due to the extra $(z-z_1)^{-1}$ factor in the integrand.  
Proceeding similarly with the other insertions, we find that \eqref{7} implies the following identity: 
\begin{align}
    &\nonumber\braket{\off{j^{\pm,\omega_1} V_{h_1,m_1}^{\omega_1}} V^{\omega_2}_{h_2,m_2} V^{\omega_3}_{h_3,m_3}}= \\
    \nonumber&(-1)^{\pm \omega_1} z_{12}^{\pm (\omega_1-\omega_2)-1}\of{\frac{z_{23}}{z_{13}}}^{\pm\omega_3}(m_2 \mp (h_2-1))\braket{V^{\omega_1}_{h_1,m_1} V^{\omega_2}_{h_2,m_2\pm1}V^{\omega_3}_{h_3,m_3}}+\\
    &\nonumber (-1)^{\pm \omega_1}z_{13}^{\pm(\omega_1 -\omega_3)-1} \of{\frac{z_{32}}{z_{12}}}^{\pm \omega_2}(m_3 \mp (h_3-1)) \braket{V^{\omega_1}_{h_1,m_1} V^{\omega_2}_{h_2,m_2}V^{\omega_3}_{h_3,m_3\pm 1}}\nonumber +\\
    & \mp (m_1 \mp ( h_1-1))\of{\frac{\omega_2}{z_{12}}+\frac{\omega_3}{z_{13}}}\braket{V^{\omega_1}_{h_1,m_1\pm 1} V^{\omega_2}_{h_2,m_2}V^{\omega_3}_{h_3,m_3}}\label{Current Correlator 2}.
\end{align}
This allows us to compute $m$-basis correlators with descendant insertions we need in terms of known primary correlators with spectral flow. 

Actually, when spectral flow is conserved, i.e.~when  $\omega_1+\omega_2+\omega_3=0$, the expression obtained in Eq.~\eqref{Current Correlator 2} can be simplified further by making use of the recursions relating the different terms on the RHS. Alternatively, we note that in this case, it turns out that 
\beq
\oint_{\cC} dz \frac{z-z_3}{z-z_1}\langle \cJ^{\pm,\omega_i}(z)V^{\omega_1}_{h_1,m_1} V^{\omega_2}_{h_2,m_2} V^{\omega_3}_{h_3,m_3} \rangle = 0. 
\label{AltCons}
\eeq
This can be used to compute the descendant correlator directly, giving 
\begin{align}
    &\nonumber\braket{\off{j^{\pm,\omega_1} V_{h_1,m_1}^{\omega_1}} V^{\omega_2}_{h_2,m_2} V^{\omega_3}_{h_3,m_3}}= 
    \mp (m_1 \mp ( h_1-1))\of{\frac{\omega_2}{z_{12}}+\frac{\omega_3\pm 1}{z_{13}}}\braket{V^{\omega_1}_{h_1,m_1\pm 1} V^{\omega_2}_{h_2,m_2}V^{\omega_3}_{h_3,m_3}} \\
    & \qquad +
    (-1)^{\pm \omega_1} z_{12}^{\pm (\omega_1-\omega_2)-1}\of{\frac{z_{23}}{z_{13}}}^{\pm\omega_3+1}(m_2 \mp (h_2-1))\braket{V^{\omega_1}_{h_1,m_1} V^{\omega_2}_{h_2,m_2\pm1}V^{\omega_3}_{h_3,m_3}}
\end{align}
An analogous expression can be obtained by inserting an extra $(z-z_2)$ power in the integrand of Eq.~\eqref{AltCons} instead of $(z-z_3)$. 

Of course, a similar computation can be carried out in the SU(2) sector of the theory. Using 
\beq
\cK^{\pm,\omega_i}(z) \equiv k^{\pm}(z)\prod_i (z-z_i)^{\pm \omega_i}\label{cKw}\,,
\eeq
leads to 
\begin{align}
    &\nonumber\braket{\off{k^{\pm,\omega_1} W_{l_1,n_1}^{\omega_1}} W^{\omega_2}_{l_2,n_2} W^{\omega_3}_{l_3,n_3}}= \\
    \nonumber&(-1)^{\pm \omega_1} z_{12}^{\pm (\omega_1-\omega_2)-1}\of{\frac{z_{23}}{z_{13}}}^{\pm\omega_3}(l_2+1\pm n_2)\braket{W^{\omega_1}_{l_1,n_1} W^{\omega_2}_{l_2,n_2\pm1}W^{\omega_3}_{l_3,n_3}}+\\
    &\nonumber (-1)^{\pm \omega_1}z_{13}^{\pm(\omega_1 -\omega_3)-1} \of{\frac{z_{32}}{z_{12}}}^{\pm \omega_2}(l_3+1\pm n_3) \braket{W^{\omega_1}_{l_1,n_1} W^{\omega_2}_{l_2n_2}W^{\omega_3}_{l_3,n_3\pm 1}}\nonumber +\\
    & \mp (l_1+1 \pm n_1)\of{\frac{\omega_2}{z_{12}}+\frac{\omega_3}{z_{13}}}\braket{W^{\omega_1}_{l_1,n_1\pm 1} W^{\omega_2}_{l_2,n_2}W^{\omega_3}_{l_3,n_3}}.
\end{align}
For $\w_1+\w_2+\w_3=0$ this takes the form 
\begin{align}
    &\nonumber\braket{\off{k^{\pm,\omega_1} W_{l_1,n_1}^{\omega_1}} W^{\omega_2}_{l_2,n_2} W^{\omega_3}_{l_3,n_3}}=  
    \mp (l_1+1\pm n_1)\of{\frac{\omega_2}{z_{12}}+\frac{\omega_3\pm 1}{z_{13}}}\braket{W^{\omega_1}_{l_1,n_1\pm 1} W^{\omega_2}_{l_2,n_2}W^{\omega_3}_{l_3,n_3}} \\
    & \qquad + 
    (-1)^{\pm \omega_1} z_{12}^{\pm (\omega_1-\omega_2)-1}\of{\frac{z_{23}}{z_{13}}}^{\pm\omega_3+1}(l_2+1\pm n_2)\braket{W^{\omega_1}_{l_1,n_1} W^{\omega_2}_{l_2,n_2\pm1}W^{\omega_3}_{l_3,n_3}}
\end{align}

\subsection{Short strings: chiral primary three-point functions}

In this section, we compute supersymmetric spectrally flowed  correlators involving short strings in the $x$-basis by using the $m$-basis results obtained above. We first consider situations where the total spectral flow charge vanishes, and then move on to general configurations.

Before continuing, let us stress that even though we have obtained closed-form expressions for $m$-basis three-point functions with arbitrary spectral flow assignments, the strategy we shall use for computing $x$-basis correlators based on these results has somewhat limited potential. It was discussed in \cite{Cagnacci:2013ufa} in the context of the bosonic SL(2,$\R$)-WZW model that if all three primary vertex operators involved in a three-point function have non-zero spectral flow, the $m$-basis techniques can not be used to obtain the $x$-basis ones for generic values of the corresponding spins $h$. 

Here we find that a similar obstruction is also present for bosonic correlators involving extra current insertions, relevant for the supersymmetric model. In both cases, the problem is that the only way to reduce the $x$-basis correlator to a known $m$-basis one is by taking limits such as $(x_1,x_2,x_3) \to (0,0,\infty)$, which are not well-defined in general. This was highlighted recently in \cite{Dei:2021xgh}. The computation of the remaining descendant correlators should follow from a non-trivial generalization of the approach used in that paper. We leave this important investigation for future work. 
In this paper, we restrict ourselves to situations involving only two spectrally flowed states. For this class of correlators, the $x$-basis fusion rules of \cite{Maldacena:2001km} reduce to the $m$-basis ones. 

\subsubsection{Spectral flow \textit{conservation}}

Let us set $\w_1 = 0$. Similarly to what was discussed in \cite{Cagnacci:2013ufa}, these correlators can only be non-zero if $\w_2 = \w_3$ or $|\w_3-\w_2| = 1$. Recall that all spectral flow charges are non-negative in the $x$-basis terminology. We first focus on the case where spectral flow is conserved and consider correlators of the form 
\beq
\braket{\cV^{(0)}_{h_1}(x_1)\cV^{\omega}_{h_2}(x_2)\cV^{\omega}_{h_3}(x_3)}. \label{VVwVw}
\eeq
Here we have chosen to write the unflowed operator in its ghost picture-zero version, given in \eqref{PicZeroUnflowed}. Then we have to compute 
\bea
&&\braket{\cV^{(0)}_{h_1}(x_1)\cV^{\omega}_{h_2}(x_2)\cV^{\omega}_{h_3}(x_3)}=\\
&&\braket{\offf{\off{j(x_1)+(1-h_1)\hat{j}(x_1)+\frac{2}{n_5}\psi(x_1)\chi_\alpha P^\alpha_{y_1,h_1-1}}V_{h_1}(x_1)W_{h_1-1}}\cV^{\omega}_{h_2}(x_2)\cV^{\omega}_{h_3}(x_3)}\nonumber
\eea
Given that both vertex $\cV^{\omega}_{h_2}$ and $\cV^{\omega}_{h_3}$ have the same SL(2,$\RR$) and SU(2) fermion number only the terms with even fermion number of $\cV^0_{h_1}$ will be non-zero. Hence, the computation of \eqref{VVwVw} involves that of the bosonic correlator 
\beq
\braket{\of{j V_{h_1}}(x_1)V^{\omega}_{h_2}(x_2)V^{\omega}_{h_3}(x_3)}.
\eeq
As argued in Sec.~\ref{sec: unflowed 3point}, we have 
\bea
&&\braket{\of{j V_{h_1}}(x_1)V^{\omega}_{h_2}(x_2)V^{\omega}_{h_3}(x_3)}=\\ &&\qquad \qquad C^{\omega}(h_i)\frac{x_{12}x_{13}}{x_{23}}\frac{z_{23}}{z_{12}z_{13}} \frac{x_{12}^{h_{\omega_3}-h_1-h_{\omega_2}}x_{23}^{h_1-h_{\omega_2}-h_{\omega_3}}x_{13}^{h_{\omega_2}-h_{\omega_3}-h_1}}{z_{12}^{\Delta_{h_1}+\Delta^{\omega}_{h_2}-\Delta^{\omega}_{h_3}}z_{23}^{\Delta^{\omega}_{h_3}+\Delta^{\omega}_{h_2}-\Delta_{h_1}}z_{13}^{\Delta_{h_1}+\Delta^{\omega}_{h_3}-\Delta^{\omega}_{h_2}}},\nonumber
\eea
where $h_{\omega_2} = h_2 + k\omega/2$ and $h_{\omega_3} = h_3 + k\omega/2$. By using the identities 
\beq
\lim_{x\rightarrow 0} V^{\omega}_{h}(x) = V^{-\omega}_{h,-h}\qqquad \lim_{x\rightarrow \infty} x^{2(h+k\omega/2)}V^{\omega}_{h}(x) = V^{\omega}_{h,h}\,,
\eeq
we obtain the relation 
\beq
\frac{\delta^2(m_1 + h_2 - h_3)C^{\omega}(h_i)}{z_{12}^{\Delta_{h_1}+\Delta^{\omega}_{h_2}-\Delta^{\omega}_{h_3}}z_{23}^{\Delta^{\omega_3}_{h_3}+\Delta^{\omega_2}_{h_2}-\Delta_{h_1}}z_{13}^{\Delta_{h_1}+\Delta^{\omega_3}_{h_3}-\Delta^{\omega_2}_{h_2}}}= \frac{z_{12}z_{13}}{z_{23}}\braket{\of{j V_{h_1}}_{h_1-1,m_1}V^{\omega}_{h_2,h_2} V^{-\omega}_{h_3,-h_3}}\label{8}
\eeq
which will allow us to compute the structure constant. Here $\of{j V_h}_{h-1,m}$ is defined as in Eq.~\eqref{jVm}. 
The three distinct terms appearing on the RHS of Eq.~\eqref{8} can be written using \eqref{Current Correlator 2}. More explicitly, we get  
\bea
\braket{\off{j^+ V_{h_1,m_1-1}}V^{\omega}_{h_2,h_2} V^{-\omega}_{h_3,-h_3}}& =&  (h_3-h_2-  h_1)\of{\frac{\omega}{z_{13}}-\frac{\omega+1}{z_{12}}}\braket{V_{h_1,m_1} V^{\omega}_{h_2,h_2}V^{-\omega}_{h_3,-h_3}}, \nn \\
\braket{\off{j^- V_{h_1,m_1+1}}V^{\omega}_{h_2,h_2} V^{-\omega}_{h_3,-h_3}}& = &  (h_3-h_2+  h_1)\of{\frac{\omega}{z_{12}}-\frac{\omega+1}{z_{13}}}\braket{V_{h_1,m_1} V^{\omega}_{h_2,h_2}V^{-\omega}_{h_3,-h_3}}, \nn \\
\braket{\off{j^{0} V_{h_1,m_1}} V^{\omega}_{h_2,h_2} V^{-\omega}_{h_3,-h_3}}&=& \of{\frac{h_2+k\omega/2}{z_{12}}-\frac{h_3+k\omega/2}{z_{13}}}\braket{V_{h_1,m_1} V^{\omega}_{h_2,h_2}V^{-\omega}_{h_3,-h_3}},
\eea
where we have set $m_1 = h_3-h_2$. This leads to 
\beq
\frac{z_{12}z_{13}}{z_{23}}\braket{\of{j V_{h_1}}_{h_1-1,m_1}V^{\omega}_{h_2,h_2} V^{-\omega}_{h_3,-h_3}} = \off{(2\omega+1)h_1-h_{\omega_2}-h_{\omega_3}}\braket{V_{h_1,m_1}V^{\omega}_{h_2,h_2} V^{-\omega}_{h_3,-h_3}}
\eeq
Given that in the $m$-basis the primary structure constants only depend on the total amount of spectral flow, we get
\beq
\frac{z_{12}z_{13}}{z_{23}}\braket{\of{j V_{h_1}}_{h_1-1,m_1}V^{\omega}_{h_2,h_2} V^{-\omega}_{h_3,-h_3}} = \off{(2\omega+1)h_1-h_{\omega_2}-h_{\omega_3}}\braket{V_{h_1,m_1}V_{h_2,h_2} V_{h_3,-h_3}}.
\eeq
The $m$-basis three-point function is derived from the $x$-basis ones by taking the limits
\bea
&\braket{V_{h_1,m_1}V_{h_2,h_2} V_{h_3,-h_3}} &= \int dx_1 x_1^{m_1+h_1-1}\lim_{x_2\rightarrow\infty,x_3\rightarrow0}x_2^{2h_2}\braket{V_{h_1}(x_1)V_{h_2}(x_2) V_{h_3}(x_3)} \nn \\
&&=\delta^2(m_1+h_2-h_3)C_H(h_i).
\eea
Comparing this with Eq. \eqref{8} we get
\beq
C^\omega(h_i) = \off{(2\omega+1)h_1-h_{\omega_2}-h_{\omega_3}}C_H(h_i),
\eeq
where $C_H(h_i)$ is the unflowed three-point function. 
In other words,  
\bea
&&\braket{\of{j V_{h_1}}(x_1)V^\omega_{h_2}(x_2)V^\omega_{h_3}(x_3)} =\label{530} \nn \\[1ex]
&&\qquad \frac{x_{12}x_{13}}{x_{23}}\frac{z_{23}}{z_{12}z_{13}}\off{(2\omega+1)h_1-h_{\omega_2}-h_{\omega_3}}\braket{V_{h_1}(x_1)V^\omega_{h_2}(x_2)V^\omega_{h_3}(x_3)} .
\eea

We also need the fermionic correlator
$
\braket{\hat{j}(x_1)\psi^{\omega}(x_2)\psi^{\omega}(x_3)}$. 
Fortunately, this can be obtained using the same methods as in the bosonic case. Indeed, we have mentioned  that $\psi^{\omega}$ are flowed primary fields of spin $\hat{h}_{\omega} = -(1+\omega)$ in the $SL(2,\RR)_{-2}$ WZW model. Hence, \eqref{AltCons} implies
\beq
\braket{\hat{j}(x_1)\psi^{\omega}(x_2)\psi^{\omega}(x_3)}=\frac{x_{12}x_{13}}{x_{23}}\frac{z_{23}}{z_{12}z_{13}}2\of{1+\omega}\braket{\psi^{\omega}(x_2)\psi^{\omega}(x_3)},
\eeq
where we have set $h_1=\Delta_1=0$.

We have now obtained all necessary ingredients for computing the supersymmetric correlator \eqref{VVwVw}. All the correlators that involve current insertions are expressed in terms of primary ones and therefore
\bea
&&\braket{\cV^{(0)}_{h_1}(x_1)\cV^{\omega}_{h_2}(x_2)\cV^{\omega}_{h_3}(x_3)}=\\
&&\frac{1}{n_5}\frac{x_{12}x_{13}}{x_{23}}\frac{z_{23}}{z_{12}z_{13}}\off{(2\omega+1)h_1 - h_{\omega_2}-h_{\omega_3}+2(1+\omega)(1-h_1)}\braket{\psi^{\omega}(x_2)\psi^{\omega}(x_3)}\times\nn\\
&&\times\braket{\chi^{\omega-1}(y_2)\chi^{\omega-1}(y_3)}\braket{V_{h_1}(x_1)V^{\omega}_{h_2}(x_2)V^{\omega}_{h_3}(x_3)}\braket{W_{h_1-1}(y_1)W^{\omega}_{h_2-1}(y_2)W^{\omega}_{h_3-1}(y_3)}\nn
\eea
Using the worldsheet and conformal symmetry to trivialize the $z$- and $x$-dependence by fixing $z_1=x_1=0$, $z_2=x_2=1$ and $z_3=x_3=\infty$ as usual, we finally obtain 
\beq
\braket{\cV^{(0)}_{h_1}(0)\cV^{\omega}_{h_2}(1)\cV^{\omega}_{h_3}(\infty)}= n_5\off{1+ H_1 + H_{\omega_2} + H_{\omega_3}}D(h_i),
\label{otravezD}
\eeq
where $H_{\omega_i} = h_i - 1 + n_5\omega_i/2$ are the spacetime weights, $D(h_i)$ is the SL(2,$\RR$) and SU(2) three-point function product given by the Eq. \eqref{SL2xSU2}, namely 
\beq
D(h_i) = \sqrt{\frac{b^2\gamma(-b^2)}{4\pi \nu }}\prod_{i=1}^3 \sqrt{B(h_i)} \equiv Q\prod_{i=1}^3 \sqrt{B(h_i)}.
\label{defCproduct}
\eeq
This generalizes the results of \cite{Giribet:2007wp} and \cite{Cardona:2009hk}. 

Non-trivial current insertions also appear in similar NSNS supersymmetric three-point  functions involving short strings. We compute them analogously, our results can be summarised as follows: 
\begin{subequations}
\bea
&\braket{\cV^{(0)}_{h_1}(0)\cV^{\omega}_{h_2}(1)\cV^{\omega}_{h_3}(\infty)}&= n_5\off{1+ H_1 + H_{\omega_2} + H_{\omega_3}}D(h_i),\\
&\braket{\cV^{(0)}_{h_1}(0)\cW^{\omega}_{h_2}(1)\cW^{\omega}_{h_3}(\infty)}& = n_5\off{1+ H_1 - H_{\omega_2} - H_{\omega_3}}D(h_i),\\
&\braket{\cW^{(0)}_{h_1}(0)\cW^{\omega}_{h_2}(1)\cW^{\omega}_{h_3}(\infty)}& = n_5\off{1- H_1 - H_{\omega_2} - H_{\omega_3}}D(h_i).
\eea
\label{3FinalSusyCorrelators}
\end{subequations}
Moreover, with the series identifications discussed in Sec.~\ref{Sec:SeriesIdenfifications} all correlators of the form 
\bea
&\braket{\cV^{(0)}_{h_1}(x_1)\cV^{\omega}_{h_2}(x_2)\cW^{\omega+1}_{-h_3}(x_3)}&\qqquad\braket{\cV^{(0)}_{h_1}(x_1)\cW^{\omega}_{-h_2}(x_2)\cW^{\omega}_{-h_3}(x_3)}, \\
&\braket{\cV^{(0)}_{h_1}(x_1)\cV^{\omega}_{h_2}(x_2)\cV^{\omega+1}_{-h_3}(x_3)}&\qqquad\braket{\cV^{(0)}_{h_1}(x_1)\cV^{\omega}_{-h_2}(x_2)\cV^{\omega}_{-h_3}(x_3)}, \\
&\braket{\cW^{(0)}_{h_1}(x_1)\cV^{\omega}_{h_2}(x_2)\cV^{\omega+1}_{-h_3}(x_3)}&\qqquad\braket{\cW^{(0)}_{h_1}(x_1)\cV^{\omega}_{-h_2}(x_2)\cV^{\omega}_{-h_3}(x_3)}, 
\eea
can be obtained from those of Eq.~\eqref{3FinalSusyCorrelators} by using the relations given in Eq.~\eqref{SusySeriesIdentification}.

\subsubsection{Normalization and matching with the symmetric orbifold CFT}
\label{sec:normalization}

Let us go back to the correlators we have obtained in \eqref{3FinalSusyCorrelators}. In this section, we describe the matching with the corresponding chiral primary three-point functions in the symmetric orbifold CFT. 

As in the unflowed case, for this,  we first need to find the correct normalization for relating worldsheet operators with their holographic counterparts. More precisely, local operators of the CFT living on the AdS$_3$ boundary are given by vertex operators such as $\Vv(x,z)$ integrated over their worldsheet insertion point $z$. Indeed, note that the worldsheet two-point functions contain a divergent factor $\delta(h_1 - h_2)$. As discussed in \cite{Maldacena:2001km,Giribet:2007wp}, this divergence is cancelled by the $z$-integrations, i.e.~by fixing the insertion points at $0$ and $\infty$ and dividing by the remaining conformal volume. However, this cancellation produces an additional finite but non-trivial multiplicative factor, which depends on the $h$ and $\w$.     

This constant factor can be obtained by using the spacetime Ward identities associated with the R-symmetry currents. It was shown in \cite{Kutasov:1999xu} that the operators
\beq
\mathbb{K}^\alpha(x,\bar{x}) = - \frac{1}{n_5 c_\nu} K^\alpha \bar{j}(\bar{x})V_1(x,\bar{x}),\quad c_\nu = \frac{\pi\gamma(1-b^2)}{\nu b^2},
\eeq
provide the worldsheet representation for the corresponding currents. Then a generic vertex operator of the form $V^\omega_{h} (x) \Phi_{\mathrm{int}}$, where $\Phi_{\mathrm{int}}$ stands for the internal, fermionic and ghost contributions must satisfy 
\beq
\braket{\mathbb{K}^{\alpha}(x_1)V^\omega_{h}(x_2)\Phi_{\mathrm{int},2}V^\omega_{h}(x_3)\Phi_{\mathrm{int},3}}= \off{\frac{q_2}{x_{12}}+\frac{q_3}{x_{13}}}\braket{V^\omega_{h}(x_2)\Phi_{\mathrm{int},2}V^\omega_{h}(x_3)\Phi_{\mathrm{int},3}},\label{62}
\eeq
where $q_3 = -q_2$ denote the corresponding R-charges. Using the methods derived in this paper, we can evaluate both sides of Eq.~\eqref{62} independently. Using \eqref{530} the LHS becomes 
\beq
\frac{-q_2}{n_5c_\nu}\frac{\xb_{12}\xb_{13}}{\xb_{23}}\left |\frac{z_{23}}{z_{12}z_{13}}\right |^2\off{2\omega+1-2h_{\omega}}\braket{V_1(x_1)V^\omega_{h}(x_2)V^\omega_{h}(x_3)}\braket{\Phi_{\mathrm{int},2}\Phi_{\mathrm{int},3}},
\eeq
with $h_\omega = h + k\omega/2$. Moreover, using Eq. \eqref{defCproduct}, we have 
\beq
\braket{V_1(x_1)V^\omega_{h}(x_2)V^\omega_{h}(x_3)} = C_H(1,h,h) = D(1,h,h) = 2Q^2B(h). 
\eeq
Hence, we find that the string two-point function differs from the spacetime one by a factor 
\beq
\off{2h_{\omega}-1-2\omega}n_5c_\nu^{-1}2Q^2B(h).
\eeq
This shows that the canonically normalized spacetime operators are given by 
\bea
&&\mathbb{V}^{\omega}_{h}(x) = \frac{\cV^{\omega}_{h}(x)}{\sqrt{2c_\nu^{-1}n_5Q^2\off{2h_{\omega}-1-2\omega}B(h)}}\\ &&\mathbb{W}^{\omega}_{h}(x) = \frac{\cW^{\omega}_{h}(x)}{\sqrt{2c_\nu^{-1}n_5Q^2\off{2h_{\omega}-1-2\omega}B(h)}}.
\label{VWnormW}
\eea
The factor $2h_{\omega}-1-2\omega$ is consistent with the unflowed result obtained in \cite{Maldacena:2001km,Dabholkar:2007ey,Gaberdiel:2007vu}. Although the generalization to the flowed case was anticipated in \cite{Maldacena:2001km,Giribet:2007wp}, to the best of our knowledge, we have presented the first formal proof available in the literature.  

Having fixed the normalization in Eq.~\eqref{VWnormW}, we obtain the following spacetime three-point functions: 
\begin{subequations}
\label{Final3ptspacetime}
\bea
&\braket{\mathbb{V}^{(0)}_{h_1}\mathbb{V}^{\omega}_{h_2}\mathbb{V}^{\omega}_{h_3}}&= \frac{1}{\sqrt{N}}\off{\frac{\of{1+ H_1 + H_{\omega_2} + H_{\omega_3}}^4}{(1+2H_1)(1+2H_{\omega_2})(1+2H_{\omega_3})}}^{1/2},\\
&\braket{\mathbb{V}^{(0)}_{h_1}\mathbb{W}^{\omega}_{h_2}\mathbb{W}^{\omega}_{h_3}}&=  \frac{1}{\sqrt{N}}\off{\frac{\of{1+ H_1 - H_{\omega_2} - H_{\omega_3}}^4}{(1+2H_1)(2H_{\omega_2}-1)(2H_{\omega_3}-1)}}^{1/2},\\
&\braket{\mathbb{W}^{(0)}_{h_1}\mathbb{W}^{\omega}_{h_2}\mathbb{W}^{\omega}_{h_3}}&=  \frac{1}{\sqrt{N}}\off{\frac{\of{1- H_1 - H_{\omega_2} - H_{\omega_3}}^4}{(2H_1-1)(2H_{\omega_2}-1)(2H_{\omega_3}-1)}}^{1/2},
\eea
\end{subequations}
where we call $H_{\omega_i}$ to the spacetime weight of each field, being $H_{\omega} = h - 1 + n_5\omega/2$ for $\mathbb{V}^{\omega}_{h}$ and $H = h +n_5\omega/2$ for $\mathbb{W}^\omega_h$, and we identify 
\beq
\frac{1}{\sqrt{N}}=\frac{g_s}{\sqrt{v_4}}\sqrt{\frac{c_\nu^3n_5}{8Q^4}}
= \frac{g_s}{\sqrt{v_4}} \sqrt{\frac{2\pi^5}{\nu \gamma(1+b^2)}}, 
\eeq
where we have inserted the factor of $g_s$ appearing in the definition of the three-point function, and the volume of the $T^4$ (which should also be included in the above normalizations). Since $g_s = \sqrt{v_4 n_5/ n_1}$ and $N=n_1 n_5$, this fixes the value of $\nu$  as in the unflowed sector \cite{Dabholkar:2007ey}\footnote{We believe there is a typo in Eq.~(4.68) in \cite{Dabholkar:2007ey} the same value for $\nu$ up to an overall $\pi^4$ factor-}.   
Our results then match exactly with the corresponding predictions from the symmetric orbifold CFT, see for instance \cite{Dabholkar:2007ey,Giribet:2007wp}.

\subsubsection{Absence of spectral flow \textit{violation}}
According to the spectral flow selection rules for the three-point function, the most general non-conserving spectrally flowed correlators with only two winding states are
\begin{align}
\begin{aligned}
&\braket{\cV^{(0)}_{h_1}(x_1)\cV^{\omega}_{h_2}(x_2)\cV^{\omega+1}_{h_3}(x_3)},\\
&\braket{\cV^{(0)}_{h_1}(x_1)\cV^{\omega}_{h_2}(x_2)\cW^{\omega+1}_{h_3}(x_3)},\\
&\braket{\cV^{(0)}_{h_1}(x_1)\cW^{\omega}_{h_2}(x_2)\cV^{\omega+1}_{h_3}(x_3)},\\
&\braket{\cV^{(0)}_{h_1}(x_1)\cW^{\omega}_{h_2}(x_2)\cW^{\omega+1}_{h_3}(x_3)},
\end{aligned}
\label{ViolatingCorrs}
\end{align}
together with similar ones with $\cW^{(0)}_{h_1}(x_1)$ instead of $\cV^{(0)}_{h_1}(x_1)$. Note that these correlators can \textit{not} be identified, via the Eq. \eqref{SusySeriesIdentification}, with those derived in the previous section. In this sense, they allow us to test for genuine spectral flow \textit{violations} in the supersymmetric model. 

As an example, we will compute the three-point function
\beq
\braket{\cV^{(0)}_{h_1}(x_1)\cV^{\omega}_{h_2}(x_2)\cV^{\omega+1}_{h_3}(x_3)},
\eeq
while the rest can be treated similarly. The only possibly non-vanishing contribution is 
\bea
&&\frac{2}{n_5}\braket{\off{\psi(x_1)\chi_\alpha P^{\alpha}_{y_1,h_1-1}V_{h_1}(x_1)W_{h_1-1}(y_2)}\cV^{\omega}_{h_2}(x_2)\cV^{\omega+1}_{h_3}(x_3)}.
\eea
Using Eq. \eqref{Py} we identify
\beq
\chi_\alpha P^{\alpha}_{y_1,l_1} = \frac{1}{2}\off{\chi(y_1)\p_{y_1}-l_1\p_{y_1}\chi(y_1)}.\label{chiPy}
\eeq
The relevant fermionic three-point functions are written as
\bea
&&\braket{\psi(x_1)\psi^{\omega}(x_2)\psi^{\omega+1}(x_3)} = \cC_\psi x_{23}^{2(\omega+1)}x_{13}^2,\\
&&\braket{\chi(y_1)\chi^{\omega-1}(y_2)\chi^{\omega}(y_3)} = \cC_\chi y_{23}^{2\omega}y_{13}^2.
\eea
By taking the appropriate limits on the $x_i$ and $y_i$ variables it is possible to identify these $x$ and $y$ basis correlators with the $m$ basis ones to determine the structure constants $\cC_\psi$ and $\cC_\chi$, giving
\bea
&&\cC_\psi =n_5\braket{\psi(x_1)e^{i(\omega+1)H_1}e^{-i(\omega+2)H_1}} = n_5^{3/2}, \\
&& \cC_\chi= n_5\braket{\chi(y_1)e^{i\omega H_2}e^{-i(\omega+1)H_2}} = n_5^{3/2}.
\eea
Then, inserting the operator \eqref{chiPy} in a fermionic three-point function gives
\bea
&&\braket{\frac{1}{2}\off{\chi(y_1)\p_{y_1}-l_1\p_{y_1}\chi(y_1)}\chi^{\omega-1}(y_2)\chi^{\omega}(y_3)}=\\
&&\hspace{5cm}\frac{1}{2}\braket{\chi(y_1)\chi^{\omega-1}(y_2)\chi^{\omega}(y_3)}\off{\p_{y_1}-2l_1y_{13}^{-1}}.\nn
\eea
From the usual $y$ dependence in the SU(2) bosonic correlator we have 
\bea
&&\off{\p_{y_1}-2l_1y_{13}^{-1}}\braket{W_{l_1}(y_1)W^\omega_{l_2}(y_2)W^{\omega+1}_{l_3}(y_3)}=\label{l_1+lw2-lw3}\\
&&\hspace{5cm}\off{l_1+l_{\omega_2}-l_{\omega_3}}\frac{y_{23}}{y_{12}y_{13}}\braket{W_{l_1}(y_1)W^\omega_{l_2}(y_2)W^{\omega+1}_{l_3}(y_3)},\nn
\eea
where $l_{\omega_i} = h_i-1+k'\omega_i/2$. To determine the spectrally flowed three-point function we take $y_2\rightarrow 0$ and $y_3 \rightarrow \infty$, so that 
\bea
&&\int dy_1 y_1^{n_1-l_1-1}\lim_{y_3\rightarrow\infty}\lim_{y_2\rightarrow 0}y_3^{-2l_{\omega_3}}\braket{W_{l_1}(y_1)W^\omega_{l_2}(y_2)W^{\omega+1}_{l_3}(y_3)}=\\
&&\hspace{5cm}\cC(l_i) \delta^2(n_1+l_{\omega_2}-l_{\omega_3})=\braket{W_{l_1,n_1}W^{-\omega}_{l_2,l_2}W^{\omega+1}_{l_3,-l_3}}.\nn
\eea
Now we can use the current defined in \eqref{cKw} and note that the $\cK^{-,\omega_i}(z,z_i)(z-z_2)$
has no poles at $z \rightarrow \infty$ for $(\omega_1,\omega_2,\omega_3) = (0,-\omega,\omega+1)$. Additionally, the extra $(z-z_2)$ power guarantees that the OPE with $W^{-\omega}_{l_2,l_2}(z_2)$ has only regular terms in $(z-z_2)$. Therefore, that 
\bea
&\oint \frac{dz}{2\pi i} \braket{ \cK^{-,\omega_i}(z,z_i)(z-z_2)W_{l_1,n_1+1}(z_1)W^{-\omega}_{l_2,l_2}(z_2)W^{\omega+1}_{l_3,-l_3}(z_3)} =& \\
&\hspace{3cm}(l_1-n_1)\of{\frac{z_{12}}{z_{13}}}^{\omega+1}\braket{W_{l_1,n_1}(z_1)W^{-\omega}_{l_2,l_2}(z_2)W^{\omega+1}_{l_3,-l_3}(z_3)}&=0.
\eea
Consequently, the correlation function can only be non-zero  if $n_1 = l_1$. This, together with the SU(2) charge conservation imposes
\beq
l_1 + l_{\omega_2} - l_{\omega_3} = 0, 
\eeq
such that the prefactor appearing in  \eqref{l_1+lw2-lw3} vanish. We conclude that 
\beq
\braket{\cV^{(0)}_{h_1}(x_1)\cV^{\omega}_{h_2}(x_2)\cV^{\omega+1}_{h_3}(x_3)} = 0.
\eeq

The same occurs for all similar non-conserving winding correlators in Eq.~\eqref{ViolatingCorrs}. In other words, we find that, for short string three-point composed by two flowed states, there is no winding violation, in the sense that all the non-zero correlators for which the total spectral flow is non-zero can be directly identified, using Eq. \eqref{SusySeriesIdentification}, with configurations were the spectral flow is  conserved.
This is consistent with the analysis of \cite{Maldacena:2001km} and \cite{Giribet:2007wp}, and also with the predictions from the holographic CFT at the symmetric orbifold point.

\subsection{Long strings and spectral flow violation}
We showed in the previous section how short-string correlators vanished for non-conserving spectral flow processes. However, this is different for interactions involving long strings. Consider the following correlator
\beq
\braket{\cV^{(0)}_{h_1}(x_1)\cV_{h_2}(x_2)\cZ^{\omega=1}_{h_1,m_1;l_1}(x_3)},
\eeq
representing the interaction between two short and a singly-wound long string. By using the explicit form of the corresponding vertex operators, we find that, given the fermionic numbers in both the SU(2) and SL(2,$\RR$) sectors, only the terms involving the currents $j(x_1)$ and $\hat{j}(x_1)$ in $\cV^{(0)}_{h_1}(x_1)$ can be non-vanishing. This leads us to the computation of the bosonic three-point function 
\bea
&&\braket{\off{j V_{h_1}}(x_1)V_{h_2}(x_2)V^{\omega=1}_{h_3,m_3}(x_3)} =\\
&&\nn \hspace{3cm}\tilde{\cC}(h_i,m_3)x_{12}^{(m_3+k/2)-(h_1-1)-h_2}x_{23}^{(h_1-1)-(m_3+k/2)-h_2}x_{13}^{h_2-(m_3+k/2)-(h_1-1)},
\eea
where here $\tilde{\cC}(h_i,m_3)$ stands for a unknown structure constant. Transforming this to the $m$-basis by taking the appropriate limits we have 
\bea
&&\oint dx_2 x_2^{m_2+h_2-1}\lim_{x_1\rightarrow0}\lim_{x_3\rightarrow\infty}x_3^{2(m_3+k\omega/2)}\braket{\off{j V_{h_1}}(x_1)V_{h_2}(x_2)V^{\omega=1}_{h_3,m_3}(x_3)}=\\
&&=\nn\tilde{\cC}(h_i,m_3)\delta^2(-(h_1-1)+m_2+m_3+k/2) = \braket{\off{j^+ V_{h_1,-h_1}}V_{h_2,m_2}V^{\omega=1}_{h_3,m_3}},
\eea
and using Eq.~\eqref{Current Correlator 2} gives
\begin{align}
    \braket{\off{j^{+,\omega_1} V_{h_1,-h_1}} V_{h_2,m_2} V^{\omega=1}_{h_3,m_3}}= &\of{\frac{z_{23}}{z_{12}z_{13}}}(m_2- (h_2-1))\braket{V_{h_1,-h_1} V_{h_2,m_2+1}V^{\omega=1}_{h_3m_3}}+\\
    &\nonumber \frac{1}{z_{13}^2} (m_3 - (h_3-1)) \braket{V_{h_1,-h_1} V_{h_2,m_2}V^{\omega=1}_{h_3,m_3+ 1}}.\nonumber
\end{align}

As was done above for the spectral flow conserving case, it is also possible to derive recursion relations between the spectral flowed violating primary correlators. Consider the operator defined in \eqref{Jw}. Given that $\sum_i \omega_i = 1$, the integral 
\beq
\oint dz \cJ^{-,\omega_i}(z,z_i)(z-z_1) = \oint dz j(z)(z-z_3)^{-1}(z-z_1)
\eeq
has no poles at infinity. The integrand also acts regularly on $V_{h_1,m_1}(z_1)$. Consequently, 
\beq
\braket{\oint dz \cJ^{-,\omega_i}(z,z_i)(z-z_1)V_{h_1,-h_1}(z_1) V_{h_2,m_2+1}(z_2)V^{\omega=1}_{h_3,m_3+ 1}(z_3)}=0, \eeq
implying 
\beq
\braket{V_{h_1,-h_1} V_{h_2,m_2+1}V^{\omega=1}_{h_3,m_3}}=-\frac{z_{12}}{z_{23}z_{13}}\frac{m_2+h_2}{m_3+h_3}\braket{V_{h_1,-h_1} V_{h_2,m_2}V^{\omega=1}_{h_3,m_3+1}}.
\label{recursionw1}
\eeq
Note that we can not do the same for  $\cJ^{+,\omega_i}(z,z_i)$ since  $\sum_i \omega_i = 1$ the behavior at infinity is different. By using \eqref{recursionw1} we obtain 
\begin{align}
    \braket{\off{j^{+,\omega_1} V_{h_1,-h_1}} V_{h_2,m_2} V^{\omega=1}_{h_3,m_3}}= &\of{\frac{z_{23}}{z_{12}z_{13}}}\cH(m_i,h_i)\braket{V_{h_1,-h_1} V_{h_2,m_2+1}V^{\omega=1}_{h_3,m_3}}, 
    \label{m-basis non conserving correlator}
\end{align}
where
\beq
\cH(m_i,h_i) = \frac{(m_2-h_2+1)(m_2+h_2)-(m_3-h_3+1)(m_3+h_3)}{m_2+h_2}.
\eeq
Since $m_2 = (h_1-1) - m_3-k/2 $ this can be  re-written as 
\beq
\cH(h_i,m_3) = \frac{(h_1-h_2-m_3-k/2)(h_1-1+h_2-m_3-k/2)-(m_3-h_3+1)(m_3+h_3)}{h_1-1+h_2-m_3-k/2}.
\eeq
The primary non-conserving spectral flow three-point function of \eqref{m-basis non conserving correlator} was derived in \cite{Maldacena:2001km}. For the $x$-basis we have
\beq
\braket{V_{h_1}(x_1) V_{h_2}(x_2)V^{\omega=1}_{h_3,m_3}(x_3)} = \cG(h_i,m_3)x_{12}^{m_3+k/2-h_1-h_2}x_{23}^{h_1-m_3-k/2-h_2}x_{13}^{h_2-m_3-k/2-h_1},
\eeq
and we can obtain the $m$-basis function from
\begin{eqnarray}
    && \braket{V_{h_1,-h_1} V_{h_2,m_2+1}V^{\omega=1}_{h_3,m_3}} =  \nn \\[1ex] 
    &&\qquad = \lim_{x_1\rightarrow 0}\lim_{x_3\rightarrow\infty}x_3^{2(m_3+k/2)}\int dx_2 x_2^{m_2+1+h_2-1} \braket{V_{h_1}(x_1) V_{h_2}(x_2)
V^{\omega=1}_{h_3,m_3}(x_3)
} \nn \\[1ex]
&&\qquad  =\cG(h_i,m_3)\delta^2(m_2+(m_3+k/2)-(h_1-1)).
\end{eqnarray}
Finally, combining all the above results we get 
\beq
\tilde{\cC}(h_i,m_3) = \cH(h_i,m_3)\cG(h_i,m_3), 
\eeq
so that the supersymmetric correlator reads
\begin{eqnarray}
&&\braket{\cV^{(0)}_{h_1}(x_1)\cV_{h_2}(x_2)\cZ^{\omega=1}_{h_3,m_3;l_3}(x_3)} =n_5\cM(H_i,h_3)\cG(H_i,h_3)C_{SU(2)}(l_i),
\end{eqnarray}
with
%
\beq
\cM(H_i,h_3) = -(H_1+H_2+H_3+1)+\frac{(H_3-h_3-\frac{n_5}{2}+1)(H_3+h_3-\frac{n_5}{2})}{H_3-H_2-H_1}.
\eeq
Where $H_1 = h_1 - 1$, $H_2 = h_2-1$ and $H_3 = m_3 + n_5/2$ are the spacestime weights of each field.
Recall that for long strings the SU(2) and SL(2,$\RR$) spins $l$ and $h$ are not related and therefore there is no cancellation scheme for the product of the bosonic three-point functions involved in this result. This is in stark contrast with the short-string correlators considered above.

Spectral flow, as a generator of new string states, is a distinctive aspect of the bosonic theory in AdS$_3$, as is the potential non-conservation of the total spectral flow charge in a correlation function. Strangely, this non-conservation of spectral flow seems somewhat elusive in the context of the  supersymmetric model, so much so that it was assumed that spectral flow non-conserving processes could even not take place, at least when discussing short strings. The results reported in this section show that the violation of the conservation of the total flow charge is possible in the supersymmetric context, although restricted to the dynamics of long strings.

\section{Discussion}

This paper focuses on three-point functions as computed from NSNS sector of the worldsheet description of type IIB superstrings propagating on  AdS$_3\times S^3 \times T^4$. Although some of them were studied in \cite{Gaberdiel:2007vu,Dabholkar:2007ey,Giribet:2007wp}, a large class of such correlators were not computed before. This is because the RNS formalism used to define the supersymmetric model conspires with the usual complications coming from the presence of spectrally flowed vertex operators, and introduces several technical complications. In particular, the picture-changing procedure needed for correlators with three or more insertions generates the appearance of extra current insertions in the bosonic SL(2,$\R$)  sector. In this context, the usual strategies for obtaining descendant correlators fail due to the highly non-trivial nature of OPEs between currents and spectrally flowed operators. These OPEs contain a set of higher-order poles whose precise form is mostly unknown \cite{Eberhardt:2019ywk}.

We have presented an explicit method for overcoming these obstacles, based on the use of the \textit{generalized} currents defined in Eq.~\eqref{genJ}. We first focused on the $m$-basis, obtaining closed-form results for all three point functions involving  SL(2,$\RR$) spectrally flowed  fields with arbitrary spectral flow charges, together with the relevant current insertions. The resulting expression, given in Eq.~\eqref{Current Correlator 2}, takes the form of a linear combination of primary correlators. This allows one to compute all $m$-basis three-point functions of the supersymmetric model. 

We then turned to $x$-basis three-point functions, best suited for the comparison with the dual holographic CFT. Following the techniques of \cite{Maldacena:2001km} and \cite{Cagnacci:2013ufa}, we have been able to use our $m$-basis results to obtain all supersymmetric three-point functions where at least one of the vertex operators is unflowed. For short strings, we considered spacetime chiral primary operators. Using the language of \cite{Maldacena:2001km,Cagnacci:2013ufa}, we showed that all the so-called spectral flow \textit{violating} correlators vanish in the supersymmetric theory, while all the spectral flow \textit{conserving} ones match exactly with the predictions of the holographic CFT at the orbifold point. In doing so, we derived the normalization appropriate for short string operators in the flowed sectors of the theory, which, to the best of our knowledge  had only been discussed heuristically so far. This significantly extends the results of \cite{Dabholkar:2007ey,Giribet:2007wp}. 

Furthermore, we have also considered interactions involving long strings, which do allow for spectral flow \textit{violation}. More precisely, we have obtained the first exact non-vanishing supersymmetric amplitude for a process that can be interpreted as the absorption/emission of a long string by the background geometry, which then increases/decreases its fundamental string charge by one unit \cite{Kim:2015gak}. 

We leave for the near future the computation of the remaining supersymmetric three-point functions in the $x$-basis  with arbitrary spectral flow charges, and the corresponding matching with the orbifold CFT predictions. For this, it will be necessary to extend the methods recently developed in \cite{Dei:2021xgh,Dei:2022pkr} (see also \cite{Iguri:2022eat}) to the relevant descendant correlators. Moreover, it would also be interesting to study non-protected three-point functions \cite{deBoer:2008ss} and to compare them with those of the proposed holographically dual theory of \cite{Eberhardt:2021vsx}. Finally, it would also be important to consider four-point functions in the supersymmetric context, where complications similar to those we have tackled in this paper appear as well \cite{Cardona:2010qf,Dei:2021yom}.

\acknowledgments 
It is a pleasure to thank Andrea Dei, Gast\'on Giribet, Carmen N\'u\~nez, Osvaldo Santill\'an and David Turton for discussions, and especially Davide Bufalini for a careful reading of the manuscript. The work of J.T.~is supported by CONICET.

\bibliographystyle{JHEP}
\bibliography{refs}

\end{document}